\documentclass{ws-ijmpd}

\usepackage{lscape}
\newcommand{\be}{\begin{equation}}
\newcommand{\ee}{\end{equation}}
\newcommand{\bea}{\begin{eqnarray}}
\newcommand{\eea}{\end{eqnarray}}

\usepackage[super,compress]{cite}
\begin{document}

\markboth{L\'opez-Corredoira \& Marmet}
{Alternative ideas in cosmology}

%
\catchline{}{}{}{}{}
%

\title{Alternative ideas in cosmology\footnote{Some parts of this text were taken from
the book {\it Fundamental Ideas in Cosmology. Scientific, philosophical and sociological critical perspectives}, to be 
published in IOP-Science by the first author\cite{Lop22}.}}

\author{Mart\'\i n L\'opez-Corredoira$^{1,2}$ and Louis Marmet$^{3}$}

\address{$^1$ Instituto de Astrof\'\i sica de Canarias,\\ 
E-38205 La Laguna, Tenerife, Spain\\
E-mail: fuego.templado@gmail.com \\
$^2$ Dep. de Astrof\'\i sica, Universidad de La Laguna,\\
E-38206 La Laguna, Tenerife, Spain;\\
$^3$ Department of Physics and Astronomy, York University,\\
 Toronto, Ontario M3J 1P3, Canada}

\maketitle

\begin{history}
\received{Day Month Year}
\revised{Day Month Year}
\comby{Managing Editor}
\end{history}

\begin{abstract}
Some remarkable examples of alternative cosmological theories are
reviewed here,
ranging from a compilation of variations on the Standard Model through the more distant quasi-steady-state cosmology, plasma cosmology, or universe models as a hypersphere, to the most exotic cases including static models. 

The present-day standard model of cosmology, $\Lambda $CDM, gives us a representation of a cosmos whose dynamics is dominated by gravity (Friedmann equations derived from general relativity) with a finite lifetime,
large scale homogeneity, 
expansion and a hot initial state, together with other elements necessary to avoid certain inconsistencies with observations (inflation, non-baryonic dark matter, dark energy, etc.). There are however some models
with characteristics that are close to those of the standard model but differing in some minor
aspects; we call these
`variations on the Standard Model'. Many of these models are indeed investigated by some mainstream cosmologists: different considerations on CP violation, inflation, number of neutrino species, quark-hadron phase transition, baryonic or non-baryonic dark-matter, dark energy, nucleosynthesis scenarios, large-scale structure formation scenarios; or major variations like a inhomogeneous universe, Cold Big Bang, varying physical constants or gravity law, zero-active mass (also called `$R_h=ct$'), Milne, and cyclical models. 

At
the most extreme distance from the standard model, the static models,  a non-cosmological redshift includes `tired-light' hypotheses, which assume that the photon loses energy owing to an intrinsic property or an interaction with matter or light as it travels some distance, or other non-standard ideas. 

Our impression is that none of the alternative models has acquired the same level of development as $\Lambda $CDM in offering explanations of available cosmological observations. One should not, however, judge any theory in terms of the number of observations that it can successfully explain (ad hoc in many cases) given the much lower level of development of the alternative ones;  but by the plausibility of its principles and its potential to fit data with future improvements of the theories. A pluralist approach to cosmology is a reasonable option when the preferred theory is still under discussion.
\end{abstract}



\section{Introduction}

Standard $\Lambda $CDM cosmology is well known among professional cosmologists. Other alternative cosmological models
are not so well known and remain ignored for most of the professionals of physics and astronomy dedicated to
study the universe as a whole. One explanation for this cognitive bias is that 
most cosmologists are quite sure that $\Lambda $CDM is the correct theory, and 
they do not need to think about possible major flaws in the basic notions
of their standard theory. They do not usually work within the framework of truly alternative cosmologies with
different fundamentals because they feel that these do not at present seriously compete with the standard model. 
These alternative models are certainly less developed because cosmologists do not work on
them. It is a vicious circle. 

Nonetheless, there is a large number of tensions and problems in the standard model \cite{Lop17c,Lop22,Per21} 
that may lead us to consider that it is not the definitive model. Therefore, one may consider other
scenarios, which might solve, at least partially, some of these caveats.

There is a rich variety of alternative ideas with intelligent proposals that merit
consideration. 
Here we will offer a sample of these alternative models; not all of them, since their number
is huge and it is impossible to mention them all
in a single article. We will not give a complete list of models, but this sample is large enough to give an
idea of which theoretical approaches are being discussed in cosmology from heterodox points of view.
Defending any particular theory against standard cosmology or globally criticizing it is not our purpose
here, although we may mention some aspects that are being debated concerning them. Here, we just review part
of the literature and offer a classification of the different models.

We will mention some of the most remarkable models in the scientific literature of recent decades
and also some  contributions of minor impact, mostly coming from professional physicists or astronomers.
There is also a vast literature produced by non-professional amateurs who try to open new routes
within the golden odyssey of cosmological model creation, but  very few of them will be mentioned here.
They are examples of curious ideas which, although not fully developed, could be the  seeds of competitive models
when they are further elaborated. To a greater or lesser extent, all of these alternative models
suffer from a lack of development in comparison with the standard $\Lambda $CDM, so the first thing
we must take into account when reading this review is that they cannot
compete with the standard model in all aspects because many of these alternative ideas are in the hands
of very few individuals---occasionally only a single individual---who
lack a collaborative environment and who  cannot produce hypotheses and ad hoc
refinement of them to fit the ever increasing deluge of observational data at the same speed as
the thousands of researchers working on the standard model. In any case, it may happen that an alternative theory might
explain certain aspects of some observational data
  better than the $\Lambda $CDM model; moreover, even if it fails to explain other types 
of observations, it might be only
a matter of time and ad hoc speculation of the existence of new unknown/dark elements for it 
to be made to fit those data too. 

All the models shown in sections \ref{.varSM} and
\ref{.majorVar} are variations based on the Lema\^itre--Gamow idea of the Big Bang idea, 
but they differ in details concerning the latter development of the theory.
But there are however other models that challenge the notions of a state of the universe with a singularity, 
unlimited density of matter-energy, or other important tenets of the standard model as proposed in the 1920s--1940s, as we show in the sections \ref{.QSSC}--\ref{.static}. 
In section \ref{.clasif}, we give a summary scheme of classification of theories according to their characteristics. 
Certainly, the proposals with the most extreme difference with respect to the standard model have also the most 
serious problems to explain the cosmological observations, as those that we summarily mention in section \ref{.caveats}, but under some refinements 
possibly some of these ideas might be the seed of future competitive models.

\section{Minor variations with respect to the standard model}
\label{.varSM}

The present-day standard model of cosmology gives us a representation of 
a cosmos whose dynamics is dominated by gravity, modelled by Friedmann-Lema\^itre-Robertson-Walker (FLRW) equations derived 
from general relativity, with a finite 
lifetime, large scales homogeneity, expansion and a hot initial state, together with 
other elements necessary to avoid certain inconsistencies with observations 
(inflation, non-baryonic dark matter, dark energy, etc.). 
In this section and the following one, we enumerate some types of
variations on this model, that is, changes with respect to 
the orthodox formulation that preserve some of the most important features of Big Bang idea.

By `minor' variations, exposed in this section, we understand the proposals that keep the fundamentals
of the standard model as stated before the 1980s and look for variations in the 
type of dark matter, the different equations of state of dark energy or even without dark energy, or the
hundreds of variations on the type of inflation or alternative proposals, or other minor details of the theory.

\subsection{Antimatter and CP violation}

\begin{figure}
\vspace{0cm}
\centering
\includegraphics[width=8.7cm]{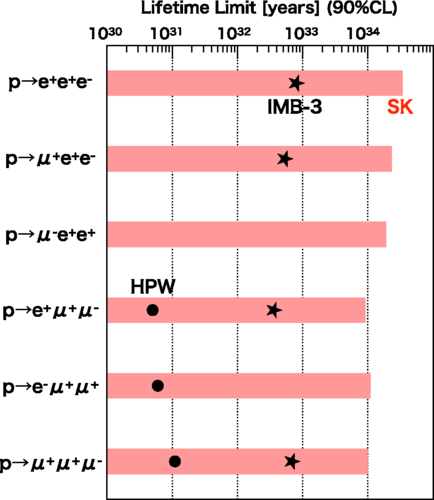}
\caption{Estimated partial lifetime lower limit for each mode of proton decay into three charged leptons.
(Reprinted from Fig. 4 of Ref. [\citen{Tan20}]).}
  \label{Fig:Tanaka2020_fig4}
\end{figure}

The topic of antimatter constitutes one of the consistency problems of the standard cosmological model. 
There are two possibilities and both are at odds with our present observations and experiments: 
\begin{enumerate}
\item If 
there were a symmetry of matter and antimatter at the beginning of the universe, where is the antimatter corresponding 
to the matter that we see now? Other galaxies cannot be made of antimatter because that would create a matter--antimatter 
boundary within the intergalactic medium that would create gamma rays that
are not observed \cite{Tau97}. 
\item If there were an asymmetry between matter and antimatter, symmetry-breaking would be needed causing other 
consequences to ensue that so far have not been successfully tested: the violation of charge-conjugation 
and parity-reversal (CP) symmetry of fundamental particles. 
This violation of CP symmetry causes most antimatter to annihilate with matter, but leaves much residual matter. 
\end{enumerate} 
 
Early models indicated that an asymmetry between matter and antimatter would also imply a finite lifetime for the
 proton. Indeed, a lifetime of around $10^{30}$ years was originally predicted by Grand Unified Theory
(GUT) \cite{Geo74} as a necessary condition for resolving the antimatter problem. 
As experiments ruled out longer and longer lifetimes, these GUT were modified to
 predict longer lifetimes \cite{Lop92}. To date, however, observations have ruled out a lifetime
up to some  $10^{34}$ yr (see Fig. \ref{Fig:Tanaka2020_fig4}).\cite{Tan20} 
There is no experimental evidence of proton decay.
 The lack of such a finite
 lifetime would rule out the baryon number non-conservation needed to overcome the $10^{11}$-fold gap between 
Big Bang baryon density predictions and observations. So this massive contradiction of prediction and
 observation still exists and  will continue to exist until we are able to observe a proton decay. 
 
Experimenters are searching for other evidences of this asymmetry. 
CP violation has been proposed for electroweak baryogenesis from the dynamical Cabibbo-Kobayashi-Maskawa matrix 
(CKM matrix, quark mixing matrix) \cite{Bru17},  CP violation in Higgs boson interactions \cite{Ber19}, and 
in leptons instead of quarks, which could generate the matter--antimatter disparity through a process called 
leptogenesis \cite{T2K20}, or other mechanisms. So far, none of these ideas has received confirmation, though they are still being actively
 researched.

\subsection{Inflation} 
\label{.inflation} 

Cosmic inflation posits 
an exponential expansion of space in the early universe. The inflationary epoch lasts from 
$10^{-36}$ s after the Big Bang singularity to some time between $10^{-33}$ and $10^{-32}$ s after it. 
The inflation paradigm explains the apparent paradox of a superluminal velocity of expansion at these short times. 
Following the inflationary period, the universe continued to expand
but the expansion was no longer accelerating \cite{Lid99}. 
It explains why the universe appears to be the same in all directions (isotropic), why the cosmic microwave 
background radiation is distributed evenly; two opposite points of the sky at the time of recombination are 
separated from each other by more than 70 times the distance that light could have travelled 
till that time. 
This smooth microwave background would be explained because inflation 
proceeds far faster than the speed of light; regions at one time in contact with 
each other, and thus at the same temperature, are blown farther 
apart from each other than the distance light could have travelled in 
the duration of the universe. 
Moreover, inflation explains why the universe is flat and the geometry nearly Euclidean 
($K=0$), and why no magnetic monopoles have been observed. 
The flatness problem is solved because the universe 
blew up to such a huge size, far bigger than the part we can observe. 
It also explains the origin of the large-scale structure of the cosmos: Quantum fluctuations in the microscopic 
inflationary region, magnified to cosmic size, become the seeds for the growth of structure in the universe. 
Different models were also constructed to explain the  origin of the initial conditions 
as seeds for the large scale structure of galaxies (e.g.\ the later-refuted baryon isocurvature 
model of
Ref. \citen{Pee89}). 
 
Some authors have argued that the inflation necessary to explain a flat universe is highly improbable \cite{Ste11}. 
If we adopt the idea of a multiverse in an eternal inflation, it has no predictive power because anything may happen.
A theory that can predict anything is a theory that predicts 
nothing. Many cosmologists see inflation as disconnected from the speculative multiverse hypothesis \cite{Cho19}, 
but even so, whether with or without multiverse, there are one thousand models of inflation \cite{Mart13}, which 
may lead us to think that almost anything is possible with inflation. 

\begin{figure}
\vspace{0cm}
\centering
\includegraphics[width=9.7cm]{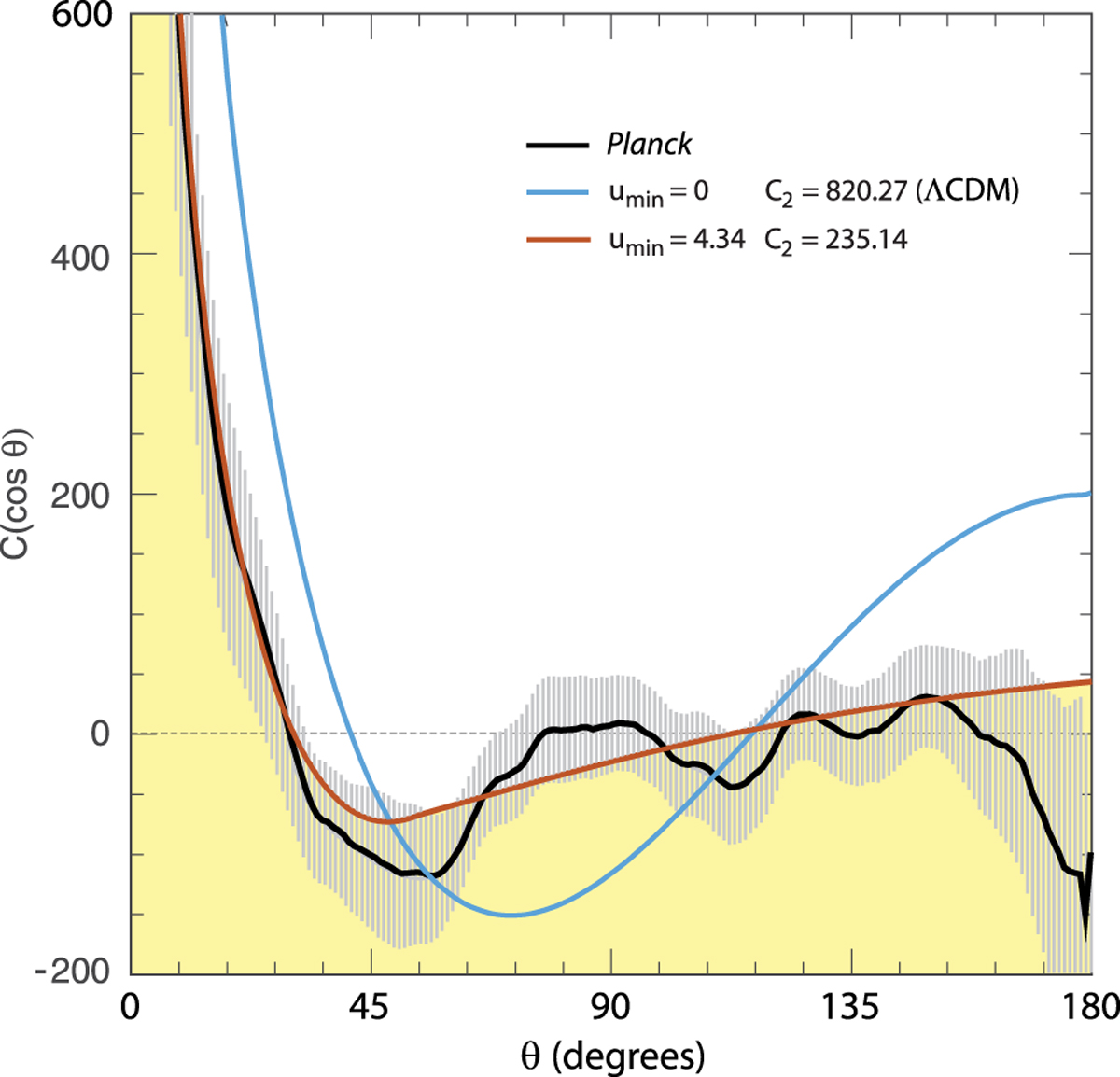}
\caption{CMBR angular correlation function measured with {\it Planck} (dark solid curve), and 
associated 1$\sigma $ errors(grey), compared with the prediction of the conventional inflationary 
$\Lambda $CDM (blue). The red curve shows the prediction of a truncated power spectrum, or a non-inflationary cosmology, with an optimized  lower limit of wave number for the fluctuation power spectrum  $u_{\rm min}= 4.34$ and quadrupole 
amplitude $C_2 = 235.14$.
(Reprinted with permission from Fig. 2 of Ref. [\citen{Mel18}]).}
 \label{Fig:Melia2018_fig2}
\end{figure}

Ref. \citen{Ilj17} argued that 
the theory is no longer scientific since it had become so all-encompassing and complicated with so many pliable 
parameters that it could be tweaked and adjusted to fit any kind of observation, and one model of inflation could be 
replaced by another as new observations appear to contradict it. 
Ref. \citen{Gut17} responded to the article by claiming that inflation is untestable because its 
predictions can be changed. Neither the attack nor the defence
puts forward
inflation as a solid theory, but merely as a speculative patch without empirical or observational support. There are even observations that disprove predictions of slow-roll 
inflation: the lack of large-angular scale correlations in
CMBR anisotropy observations is in considerable conflict with the basic inflationary paradigm \cite{Mel18} (see Fig. \ref{Fig:Melia2018_fig2}).
There is no global consensus among cosmologists concerning the reality of inflation, and 
we are far from having irrefutable proofs that keep it in line with solid evidence. 
This motivates multiple variations of the types of inflation, or even proposals without inflation
\cite{Pat08,Bra08}.

\subsection{Dark energy variations}

\begin{figure}
\vspace{0cm}
\centering
\includegraphics[width=11.7cm]{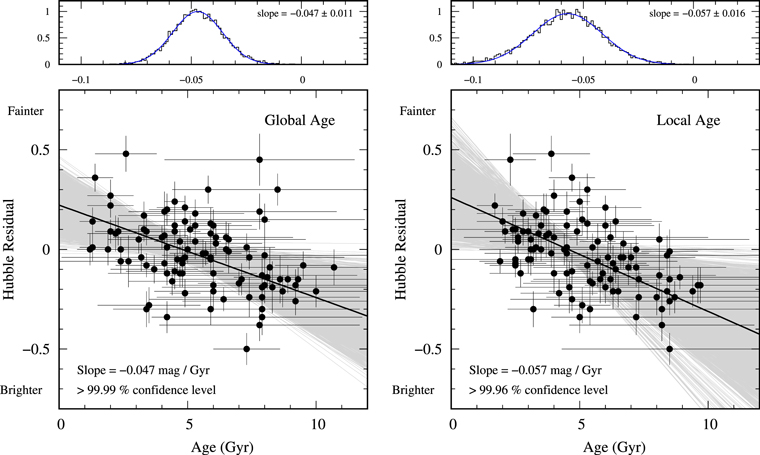}
\caption{The strong correlation between population age and Hubble Residual for host galaxies of all morphological types. The left panel is for the mass-weighted global ages of 102 host galaxies. The solid line is the best regression fit obtained from MCMC posterior sampling method, which shows a very significant (4.3$\sigma $) correlation between age and Hubble Residual with the slope. The right panel is for the local age of the environment around the site of SN in a host galaxy. Again, a similarly significant (3.6$\sigma $) correlation is obtained with an even steeper slope ($\sim $0.057 mag Gyr$^{-1}$).(Reprinted by permission of the AAS from Fig. 1 of Ref. [\citen{Lee20}]).}
    \label{Fig:Lee2020_fig1}
\end{figure}

As an alternative to dark energy it is suggested that an observational bias explains the appearance of dark
energy.\cite{Nie16,Col19}  Using a much bigger database of supernovae, statistical tests show the data are
quite consistent with a constant rate of expansion.  Observations reveal a ``bulk flow'' in the local Universe
which is faster and extends to much larger scales than are expected around a typical observer in the standard
$\Lambda$CDM cosmology.  Thus the cosmic acceleration deduced from supernovae may be an artefact of our being
non-Copernican observers, rather than evidence for a dominant component of `dark energy' in the Universe.
It was also suggested to be due to 
intergalactic dust \cite{Mil15}, or metallicity evolution of supernovae. 
There is indeed a significant correlation between SNe Ia luminosity 
and stellar population age, which causes a 
serious systematic bias with look-back time able to mimic the 
dark energy effect \cite{Kan19,Lee20} (see Fig. \ref{Fig:Lee2020_fig1}). Therefore, 
old supernovae might be intrinsically fainter than the local ones 
and the cosmological constant would not be needed.

Admitting the dark energy, there are also variations with respect to the interpretation as
cosmological constant $\Lambda$: Quintessence models of dark energy propose that
the observed acceleration of the scale factor is caused by a dynamical field.\cite{Cal98}  The equation of
state of this inhomogeneous component is different from baryons, neutrinos, dark matter, or radiation. Unlike
the cosmological constant $\Lambda$, it evolves dynamically and develops fluctuations, leaving a distinctive
imprint on the microwave background anisotropy and mass power spectrum.
Other authors play with the dark energy equation of state parameter (instead of the
standard value for a cosmological constant of $w=-1$) and its 
evolution.\cite{Tri17,Ten21}

\subsection{Scenarios without non-baryonic cold dark matter with standard gravity} 
\label{.nonCDM} 
 
Some dynamical problems  in which dark matter has been claimed as necessary can indeed be 
solved without non-baryonic dark matter: galactic stability \cite{Too81} 
or warp creation \cite{Lop02b}, for instance. Velocities in galaxy pairs and satellites 
might also measure the mass of the intergalactic 
medium filling the space between the members of the 
pairs rather than the mass of dark haloes 
supposedly associated with the galaxies \cite{Lop99,Lop02b}. 
 
Rotation curves in spiral galaxies 
can be explained within standard gravity with magnetic fields, 
non-circular orbits in the outer disc \cite{Ben17}, 
or types of dark matter different from non-baryonic cold dark matter 
proposed in the standard model \cite{Bat00}. 

There are also proposals stating that the dark matter necessary to solve many problems may be baryonic by  simply placing
baryonic dark matter in the outer disc \cite{Fen15,Sip21}. Other, more exotic, proposals include 
positively charged, baryonic  particles (protons and helium nuclei) \cite{Dre05}, which 
are massive and weakly interacting, but only when moving at relativistic velocities.
Simple composite systems include nucleons but 
are still bound together by comparable electric and magnetic forces \cite{May12}, making up a three-body
system (`tresinos') or four-body system (`quatrinos'), antiparticles, which have negative 
gravitational charge \cite{Haj14}. Then, there is the proposal, among others, of a pressure-free fluid in general relativity \cite{Coo05}. 
Nonetheless, these proposals have their own set of problems when trying to explain large-scale structure
or the Cosmic Microwave Background Radiation (CMBR). 
 
\subsection{Nucleosynthesis variations}
 
The core of Big Bang nucleosynthesis calculations is the Boltzmann equation in an expanding universe.
There are several parameters on which the equation depends:
the neutron half-life, the number of neutrinos and the baryon density. 

The neutron half-life is well known, so there is no discussion on it. The Standard Model of particle physics predicts the existence of three species of neutrinos, but different numbers have also been proposed to solve the inconsistencies. Curiously, a much better fit between primordial nucleosynthesis 
and observations is obtained
when the number of neutrino species is two instead of the three 
predicted by the standard model for particles \cite{Hat95}. 

The baryon density is perhaps the parameter with larger number of disputes.
In order to get the observed abundances of light elements with primordial nucleosynthesis, 
a baryon fraction is predicted for the $\Lambda $CDM model. However, this number does not fit 
the observations---at least not directly. 
Less than 50\% of the baryons predicted 
at low redshift have been found in some way in galaxies or 
the intergalactic medium, and the galaxies 
are observed to have a significantly smaller baryon fraction relative to the cosmic 
average \cite{And10,McG10}.  
To solve it, it was proposed that in the quark--hadron transition 
stage in the very early universe \cite{Oli91} there could emerge some regions 
which have more protons and with more neutrons than the near-equality 
assumed in the thermodynamic equilibrium value of the standard model. Nonetheless, the cosmological consequences of a quark-hadron phase transition no longer 
hold and it is still unclear how baryons (not hadrons) could form at 
that cosmological transition \cite{Bon16}.

\subsection{Large--scale structure formation variations}

\begin{figure}
\vspace{0cm}
\centering
\includegraphics[width=11.7cm]{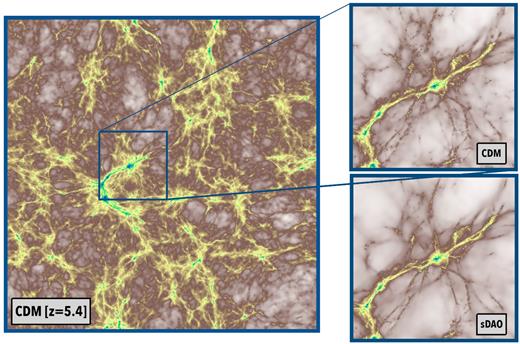}
\caption{Images of the gas density at $z=5.4$ obtained from hydrodynamical simulations by Ref. [\citen{Bos19}]. On the left, the simulation box along the $z$-axis in a projection of comoving dimensions (20x20x4) $h^{-1}$ Mpc is projected. The smaller panels zoom into a region centred on the most massive halo at this redshift in a window of size (4x4x2) $h^{-1}$ Mpc in Cold Dark Matter (CDM; upper right) and the equivalent region in {\it strong dark acoustic oscillations} (sDAO; lower right), a type
of interacting dark matter, resulting in a smoother matter distribution than the CDM volume at the same epoch, although with differences hard to discern on these scales. 
(Reprinted by permission of the AAS from Fig. 2 of Ref. [\citen{Bos19}]).}
    \label{Fig:Bose2019_fig2}
\end{figure}

In the cold dark matter (CDM) theory, the structure of the large-scale distribution of galaxies grows
hierarchically: galaxies being formed in continuous episodes of accretion and merging, with small objects
collapsing under self-gravity first and then merging in a continuous hierarchy to form larger and more
massive objects. These semi-analytical $\Lambda $CDM models claim that very massive galaxies were formed
much later than small galaxies \cite{Guo11}. It is the opposite of the hot dark matter (HDM) paradigm, which
was more commonly used in the early 1980s, where structure does not form hierarchically (bottom-up), but
forms by fragmentation (top-down), with the largest superclusters forming first in flat pancake-like sheets
and subsequently fragmenting into smaller pieces that constitute the galaxies. There are also models that
are a mixture of cold and hot dark matter, called warm dark matter (WDM), or the interacting dark matter
models, which both result in a cut-off in the linear power spectrum, that have been competing with CDM in
recent years (see, for instance, Fig. \ref{Fig:Bose2019_fig2}).\cite{Bod01,Bos19}

A monolithic scenario within CDM opposite to the hierarchical scenario, with galaxies all forming at once, is one of the minor variations available in the literature. Speculative solutions have also been proposed in terms of either a `downsizing' scenario of galaxy formation or a mass function variation with redshift \cite{Man15}. 

Indeed, current CDM models predict the existence 
of dark matter haloes for each 
galaxy whose density profile falls approximately as $r^{-2}$, although the 
original idea \cite{Whi78} concerning hierarchical structures with 
CDM that gave birth to the present models was that the dark matter was distributed 
without internal substructure, more like a halo with galaxies than 
galaxies with a halo, such that, for instance, the Milky Way, Andromeda, and other galaxies 
of the Local Group are all embedded in the same common halo. The intergalactic matter in this case is not empty 
but contains the greater part of the mass of groups or clusters of galaxies \cite{Bat00,Lop99,Lop02b}.

\section{Major variations with respect to the standard model}
\label{.majorVar}

As major variations, we will refer here to the changes with respect to the standard model that keep most of the essential points
but alter at least one of them: either homogeneity at large-scale, the hypothesis of high temperature at the beginning of the universe,
constants of physics that are rendered into varying parameters, some FLRW solutions away from the original proposals, and cyclical universes are
some of these models.

\subsection{Inhomogeneous universe}

In this variation, the density distribution of the universe is not homogeneous on very large scales. It may
obey a fractal distribution \cite{Bar94,Gab05}, when the  mass  within a sphere of radius $R$ is not
proportional to $R^3$ for large enough $R$ (in the regime in which there should be homogeneity), but
proportional to $R^D$ with a fractal dimension $D<3$. There is little theoretical background
to support a cosmology of these characteristics, but  some observations 
may point in this direction. 

Another idea stems from
timescape cosmology \cite{Wil09}, in which apparent
cosmic acceleration is an effect related to the calibration of clocks and rods of observers in bound systems relative 
to volume-average observers in an inhomogeneous geometry in ordinary general relativity. Back reactions have caused time 
to run more slowly or, in voids, more quickly, thus producing the illusion that supernovae are farther away than they
 are and also implying that the expansion of the universe is in fact slowing down.
An inhomogeneous isotropic universe described by a
Lema\^itre--Tolman--Bondi solution of Einstein's field equations can also provide
a positive acceleration of the expansion without dark energy \cite{Rom07}.

\subsection{Cold Big Bang} 

This theory evolved from the 1960s \cite{Lay90}.
Rather than a very high temperature at the beginning of the universe with later progressive
cooling, the universe starts with $T=0$ K. Explanations are offered for
the origin of the light elements in primordial and/or stellar nucleosynthesis  \cite{Agu99}, the cosmic microwave 
background radiation \cite{Agu00} in terms
of thermalization by intergalactic particles---a mixture of carbon/silicate dust and iron or carbon whiskers---of stellar 
radiation originating in Population III (further details of this idea are given in \S \ref{.QSSC}), and
other phenomena explained by the standard Hot Big Bang.

\subsection{Variations or oscillations of physical constants}

Some models propose variations or oscillations of physical constants  ($c$, $G$, $h$, the fine structure constant
$\alpha $, or others) with time or distance. For instance, Ref. \citen{Gol92} proposes variable $G$
with distance, with which they explain the apparent variation of $\Omega $ with the scale with no
need of dark matter; or a temporal variation of $G$ based on the Large Number Hypothesis
which states that large dimensionless numbers are connected with the present epoch and vary
with time.\cite{Dir74}  Other cosmological models propose a
variable speed of light $c$ over time \cite{Alb99,Sho04,Ell07} or other variables \cite{Unz15},
or several physical constants varying at the same time \cite{Gup20}. The consequences differ according to the
models. They preserve the basic aspects of the standard model while keeping expansion and finite time since the beginning of the universe, 
but they differ with regard to a number of characteristics \cite{Goh17}.

\subsection{Modifications of the gravity law} 

These theories not only change $G$, but also the gravity force equation.
There are several tens of these alternative theories which we will not mention here. They can be
found in other reviews \cite{Cli12}. 

The most popular alternative gravity theory is the modification of gravity law proposed in `Modified Newtonian Dynamics' 
(MOND) \cite{San02,Fam12,San15,Mer20}, which modifies the Newtonian law for accelerations lower than 
$\sim a_0\approx 1\times 10^{-10}$ m/s$^2$. 
This acceleration scale $a_0$ defines the variation with respect to Newton's law necessary to fit the rotation 
curves. Its value is very similar in all galaxies and it has been interpreted 
as a possible sign of confirmation of MOND \cite{Lel17}. 

MOND was in principle a phenomenological approach. Its proponents attempted to incorporate elements that
make it compatible with more general gravitation theories; for example, A QUAdratic Lagrangian theory
(AQUAL) \cite{Bek84} or Quasi linear approximation of MOND (QMOND) \cite{Mil10}, which expanded MOND to
preserve the conservation of momentum, angular momentum, and energy, and follow the weak equivalence
principle.

\begin{figure}
\vspace{0cm}
\centering
\includegraphics[width=9.7cm]{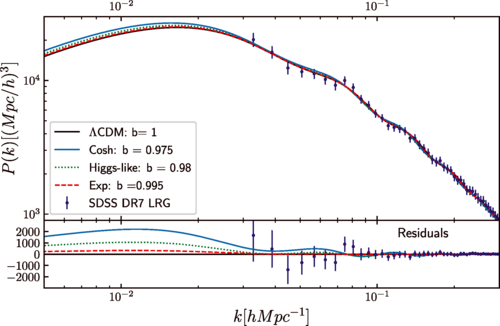}
\caption{The linear matter power spectrum $P(k)$ for $\Lambda $CDM and for different models of relativistic MOND \cite{Sko21}, showing excellent fits to the Sloan Digital Sky Survey (SDSS) data release 7 (DR7) luminous red galaxies.
(Reprinted with permission of American Phys. Soc. from Fig. 2 of Ref. [\citen{Sko21}]).}
    \label{Fig:Skordis2021_fig2}
\end{figure}

Several relativistic versions of MOND exist.  A relativistic gravitation theory of MOND was developed under
the name  Tensor-Vector-Scalar (TeVeS) \cite{Bek04}, which also tried to provide consistency with certain
cosmological observations, including gravitational lensing. In MOND/TeVeS, there are different cosmological
views. Ref. \citen{Fel84} does not accept that a MOND-cosmology might be possible, stating that a
quasi-Newtonian calculation adapted from Newtonian cosmology suggests that a MOND universe will recollapse
and/or fail to satisfy the cosmological principle of a homogeneous universe. Other authors claim that a
MOND-cosmology can be built \cite{San98,San06} that results in a uniform expansion and homogeneity on the
horizon scale consistent with MOND-dominated non-uniform expansion and the development of inhomogeneities on
scales out to a substantial fraction of the Hubble radius. Primordial nucleosynthesis, with its concomitant 
thermal and dynamical history of the universe, is identical to that of the standard cosmological model until
matter dominates the energy density of the universe, a moment in which the MOND cosmology diverges from that
of the standard model. The most recent version RMOND\cite{Sko21} claims to successfully reproduce key
cosmological observables such as the CMBR and the matter power spectrum of the universe (see Fig.
\ref{Fig:Skordis2021_fig2}).

Tentative detections of a departure from Newtonian gravity as  predicted by MOND was  claimed, for example,  
for wide binaries (binary stars with distances of several thousands of AU), similar to the expectations of
 modified gravity theories \cite{Her12}, although this was done by comparing only the dispersion of radial
 heliocentric velocities, and the gravitational effects 
of the surrounding areas are not clear. Moreover, the distribution of separations of wide binaries and their evolution \cite{Pov04,Jia10b} 
should also have an effect  that should 
be considered carefully in this test. 
Nevertheless, the successes of MOND and its relativistic version are mostly limited 
to galactic scales, the model has a missing mass problem for clusters of galaxies\cite{Fam12}
and the relativistic extensions need ad-hoc assumptions to describe the necessary
phenomenological facts.
  
A different alternative gravity theory with a certain impact is a scalar-tensor-vector one
known as modified gravity 
(MOG, Ref. \citen{Mof06}). Another family of theories is the $f(R)$ gravity group, which modify general relativity by defining a 
different function of the Ricci scalar \cite{Sot10}. Other proposals include the dependence of space-time on curvature  in a non-metric theory of 
gravity \cite{Kra08}, or an interpretation of Mach's principle in which the rotational reference frames for stars in 
galactic orbits have a relationship to the rotating matter
in the local galaxy and/or  distant galaxies \cite{Ann16}.   
There are many others \cite{Cli12}: Einstein-ether theory, bimetric or general higher-order theories, 
Ho\v{r}ava-Lifschitz gravity, Galileons, Ghost Condensates, and models of extra dimensions, including Kaluza-Klein, 
Randall-Sundrum, Dvali-Gabadadze-Porrati model 4D gravity on a brane\footnote{In string theory and related theories 
such as supergravity theories, a brane is a physical element that generalizes the notion of a point particle to higher 
dimensions. Branes are dynamical objects that can propagate through spacetime according to physical laws of quantum 
mechanics.} in 5D Minkowski space, or higher co-dimension braneworlds, Weyl conformal gravity (invariant under Weyl transformations \cite{Jac14}), etc. 
The cosmological implications of these changes in gravity laws are important, but in most cases they have not
yet been developed.

\subsection{Other Friedmann-Lema\^itre-Robertson-Walker (FLRW) solutions}

Instead of the particular solutions of $\Lambda $CDM with fixed parameters $\Omega _m\approx 0.3$,
$\Omega _\Lambda \approx 0.7$, the FLRW equations may have other solutions. Here we have a couple of examples:

One remarkable case that has generated a large number of papers in recent years is the Zero-active mass condition, also called 
$R_h=c\,t$. This model was firstly proposed under the name of an `Ur theory', which relates cosmology to particle 
physics and quantum theory \cite{Gor88}, or a special (flat) case of an eternal coasting model \cite{Joh96}, consistent 
with a scale factor proportional to cosmic time ($a(t)\propto t$) or 
equivalently an active mass $\rho +3\frac{p}{c^2}$ equal to zero at all times, which, together with the ansatz $\Lambda =0$, makes 
the acceleration of the expansion equal to zero for all times. The density is fitted to keep the universe flat.

The most important researcher defending
this model nowadays is Fulvio Melia, who has produced several tens of papers with theoretical and 
observational support for the model \cite{Mel12,Mel15,Mel17,Mel19}. 
There is the coincidence that now the deceleration of the Hubble--Lema\^itre flow
is compensated by the acceleration of the dark energy;
the average acceleration throughout the history of the universe is almost null \cite{Mel12} 
and the size of the universe is such as if there were constant expansion. This coincidence supports
the constant expansion ratio posited as $R_h=c\,t$.
According to Melia,
this model fits the data pretty well  where $\Lambda $CDM does.  Moreover, it offers some further advantages
at high $z$, where the standard model has some difficulties in concealing the existence of objects
that usually need a long time to be formed in a very young universe that does not allow time for such evolution.  
$R_h=c\,t$ solves the problem because the age of the universe at redshift $z$, $t(z)$, is much greater in this model than with the 
standard $\Lambda $CDM. The Big Bang would have happened $H_0^{-1}=14.57$ Gyr ago (for $H_0=67.4$ km/s/Mpc; Ref. \citen{Pla20}), longer 
than the 13.79 Gyr for $\Lambda CDM$ with the same Hubble--Lema\^itre constant. There would have been no inflation.
Interestingly, an even more remarkable difference from the standard model is the
prediction that CMBR is formed at $z\approx 16$ by dust rethermalization of Population III stellar light
\cite{Mel20}, although with a major difficulty in justifying a perfect blackbody shape for that radiation, since
we do not know of any kind of dust that produces such a flux shape.
This very different explanation of CMBR makes a major difference
with the standard model.

Melia also points out that the $R_h=ct$ model has a higher probability of 
being correct than $\Lambda $CDM, even with the same quality of fits, based on the fact that $R_h=ct$
has no free parameters in the fits, whereas $\Lambda $CDM has. However, we must bear in mind that the
election of the $\rho +3\frac{p}{c^2}=0$ $\forall t$ condition is posited because, a priori, we know it gives similar results to the standard models with $\Omega _m=0.3$ 
in many cosmological tests at low and intermediate redshift, and non-acceleration of the expansion at present, so there is an implicit choice of parameters.
Melia thinks that this choice is not arbitrary and  derives it from general relativity for FLRW metrics, but this interpretation of general relativity is not supported by other specialists in gravitation  (Ref. \citen{Kim16}; a paper replied to by Melia in Ref. \citen{Mel17}).

Another example of an FLRW model is Milne Cosmology, which is obtained by demanding that the energy density, pressure, and 
cosmological constant are all equal to zero, and that the spatial curvature is negative. Indeed, this can also be considered as a Zero 
active model, but with the particular case of $\rho =0$. This strictly null density of matter is
unrealistic since stars, gas and other components of the universe have mass and are real, but
might possibly consider the density to be low enough to be considered close to zero, in comparison
with the critical density in the standard model. Also, a symmetric Milne universe \cite{Ben08}, 
composed of matter with positive and negative mass in equal quantities would have net $\rho =0$.
From these assumptions and the FLRW equations, it follows that the scale factor must depend
linearly on the time coordinate, $a(t)\propto t$, as in all zero-active mass cases. The model has
no expansion of space, but the mathematical equivalence of the zero energy density version of the
FLRW metric to Milne's model implies that a full general relativistic treatment using Milne's
assumptions would result in an increasing scale factor and an associated metric expansion of space
\cite{Cha15}. The Milne Cosmology does not need dark matter or dark energy; moreover, the model
evades horizon, flatness, and cosmological constant problems affecting the standard cosmology,
which requires inflation to solve it; it also fits some cosmological data \cite{Ben08,Vis13}.

\subsection{Cyclical universes}
\label{.cyclical}

In the `Conformal cyclic cosmology' model \cite{Pen10}, based in the framework of general relativity, the
universe iterates through infinite cycles, with the future timelike infinity of each previous iteration
being identified with the Big Bang singularity of the next. So the Big Bang (although without inflation)
applies to our present universe, but it is speculated that many other Big Bangs happened previously and will
happen after ours. The Big Bang of each is taken to be a smooth conformal continuation of the remote future
of the previous one via an infinite conformal rescaling; there is no collapsing phase. The second law of
thermodynamics, with the curious nature of its origin, is automatically incorporated, where Hawking
evaporation of black holes provides a key ingredient.

Another variation of the cyclical model proposed by Ref. \citen{Ste07b} is an endless universe without beginning or end,
an endless sequence of epochs that starts with a `Big Bang' (again without inflation)
and ends in a `Big Crunch' (collapse of the universe due to an excess of mass-energy of the critical density in 
a closed universe). Although the model is motivated by M-theory,\footnote{M-theory unifies all consistent versions of superstring 
theory.} branes, and extra dimensions, the scenario can be described almost entirely in terms of conventional 4D field theory and 4D cosmology, 
with a continual cycle of expansion and contraction as parallel universes (or `branes') collide. 

In the `Dynamic universe' \cite{Sun20}, the universe is also described as
a contraction-expansion cycle from infinity in the past to infinity in the future or in repeated cycles. 
In the contraction phase, mass in space gets its energy of motion from its own gravitation
and releases it back to gravity in the ongoing expansion phase. 
Phenomena that general relativity theory explains in terms of modified space-time metrics
are explained as consequences of their different energy states.  The buildup of local structures in space
converts part of the momentum in the fourth dimension into momentum in space via local bending of space.
Galactic space in the DU appears in Euclidean geometry, and the magnitudes of high redshift supernovae
are explained without assumptions of dark energy or accelerating expansion.

\section{Quasi-Steady State Cosmology}
\label{.QSSC}

All the previous models are variations based on the Lema\^itre--Gamow idea.
But there are however other models that challenge the notions of a state of the universe with a singularity, 
unlimited density of matter-energy, or other important tenets of the standard model as proposed in the 1920s--1940s.
Among the most developed models, with several professional researchers working on it over several decades, 
perhaps the alternative hypothesis of highest impact during the last century is the `Quasi-Steady State Cosmology' (QSSC), which was indeed first called the `Steady State Cosmology' when it was a
cosmological model competing at the same level of importance and impact with the Big Bang hypothesis.
It was developed and defended for more than sixty years, and even today we cannot declare it dead,
although its main supporters have passed away or retired, and the new generations of cosmologists
no longer work on it. 

This model is indeed something beyond a small variation on Lema\^itre-Gamow ideas, it is a radically different view
in which there is no beginning of the universe, but instead
an eternal cycle of matter creation. The
difference with the cyclical universes of \S \ref{.cyclical} is that there is no singularity of state
of infinite or unlimited density, and the moment of `creation' is not unique, but
continuous and constant (in the 
Steady State version) or oscillating (in the Quasi-Steady State version).

Fred Hoyle, and independently Hermann Bondi and Thomas Gold, proposed in 1948 
the hypothesis of the Steady State \cite{Hoy48,Bon48} in which, contrary to the Big Bang 
approach, there was no beginning of the universe.\footnote{Indeed, prior to Hoyle, Bondi, or Gold,
Albert Einstein attempted to construct a Steady-State model of the universe, as shown in one of his
unpublished manuscripts \cite{ORa14}. The manuscript, which appears to have been written in early
1931, demonstrates that Einstein once explored a cosmic model in which the mean density of matter
in an expanding universe is maintained constant by the continuous formation of matter from empty
space. Einstein's Steady-State model contained a fundamental flaw and this was possibly the reason
why he abandoned this line of research \cite{ORa14}.}
The universe is expanding, it is
eternal, and the homogeneous distribution of matter is being created at a rate of $\sim 10^{-24}$ 
baryons/cm$^3$/s owing to the existence of a putative ubiquitous C-field of matter creation, 
instead of the unique moment of creation in
the Big Bang.
The perfect cosmological principle of a universe being observationally the same  from 
anywhere in space and at any time is maintained in this model, whereas the standard model only gives a 
cosmological principle in space but not in time. There is no evolution. The universe remains always 
the same. 
The matter distribution is homogeneous
and the redshift is caused by the expansion of the space with a scale of
$S(t)\propto e^{H_0t}$. Newly created matter forms new galaxies that substitute those that are swept 
away by the expansion.

During the 1950s, both the Big Bang and Steady-State theories held their ground. While there were attempts 
to explain the abundances of the chemical elements with Gamow et al.'s theory,  the Steady State Theory
also provided plausible explanations.
The abundances of the light elements 
(helium, lithium, deuterium, and others) were explained in terms
of stellar nucleosynthesis and collision with cosmic rays in the remote past of the universe,
as proposed by Margaret Burbidge and collaborators\cite{Bur57}.
The heaviest elements could also be explained in terms of stellar rather
than primordial nucleosynthesis, and the defenders of Big Bang in the end also had to adopt the
stellar nucleosynthesis of Burbidge et al.\ for the heavy elements.

Nonetheless, the Steady State theory would lose competitiveness by the mid-sixties,
because it could not explain certain observational facts. It could not explain why the galaxies
were younger at higher redshift. Neither could it explain the excess of radio sources at large
distances \cite{Ryl61}, or the distribution of quasars. Most importantly,
it did not explain the CMBR as interpreted in cosmological terms in 1965. This strongly favoured the Big Bang theory.

\begin{figure}
\vspace{0cm}
\centering
\includegraphics[width=13cm]{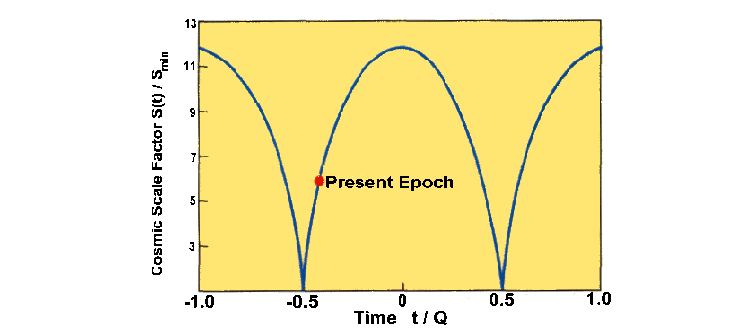}
\caption{Oscillation of the cosmic scale factor $S(t)$ in the quasi-steady-state cosmology, for the
case in which the matter creation rate is assumed to be small. The ratio of $S$ to its minimum
value is plotted against time normalized to the characteristic oscillation time,
$Q=10^{11}$ yr. The red dot marks the present epoch.
(Reprinted from https://casswww.ucsd.edu/archive/personal/gburbidge.html).}
    \label{Fig:Burbidge_UCSD}
\end{figure}

In 1993--94, Fred Hoyle, Geoffrey Burbidge, and Jayant Narlikar published a modification of the model 
that was called the `Quasi-Steady State' theory \cite{Hoy93,Hoy94,Hoy00,Nar07}. 
The main modification consisted in positing an oscillatory expansion apart from the exponential term
(see Fig. \ref{Fig:Burbidge_UCSD}):
\begin{equation}
S(t)\propto e^{t/P}[1+\eta \cos(2\pi \theta(t)/Q)] 
,\end{equation}
with a long time scale of expansion of $P\sim 10^{12}$ years, $\theta (t)\sim t$. 
The exponential factor had already been introduced in the first version of the Steady State model to keep
$\frac{\dot{S}}{S}=$constant and consequently maintain a constant density of matter by invoking the continuous creation of matter. 
The new term here is the sinusoidal oscillation with period $Q$.
The exact value of the parameter $Q$ and $\eta $ would be determined from Hubble--Lema\^itre's constant, the age of 
globular clusters, and the maximum observed redshift in the galaxies. 
The time since the last maximum of $S(t)$ is $0.85Q$.
The time since the last minimum of $S(t)$ is 14 Gyr.
With the parameters given in the original version of the  QSSC theory, 
the maximum observable redshift should be around 5, although it would be increased later as the parameters changed
to adapt to new observations.
The creation of matter is confined to epochs with minimum $S(t)$ rather than being continuous.
These creation events involve Planck particles and eventually
make hydrogen gas plus the lightest elements, deuterium and the two
isotopes of helium. Since the overall time scale is very long, many generations of stars
(and galaxies) will evolve and die.

With the introduction of the Quasi-Steady variation,
some of the problems that affected the original theory of 1948 were solved.
This explained why there are younger galaxies at higher redshift, the problem of the 
radio sources, the distribution of quasars (with lower density for redshifts lower than 2.5), and
the formation of large-scale structure (clusters, voids, filaments) \cite{Nay99}.
Ref. \citen{Wri95} complained that Hoyle et al. had not solved the problem of the radio sources
completely, but Ref. \citen{Hoy95} later replied that the question might be solved with a change of parameters. 

The CMBR and its blackbody spectrum would be explained
as the effect of the thermalization of radiation emitted by stars of the last cycle
$P/3$ due to absorption and re-emission that produce iron needle-shaped particles (`whiskers')
in the intergalactic medium. Because of the long distances travelled by the CMBR photons in the maxima
of the oscillation and  the thermalization that occurs at each minimum, there is
no accumulation of anisotropies from one cycle to another.
Only the fluctuations of the last minimum survive, which gives fluctuations of temperature
comparable to the observed $\Delta T/T \sim 5 \times 10^{-6}$. 
First, the carbon needles thermalize the visible light from the stars, giving rise to 
far-infrared photons at $z\sim 5$, thus maintaining the isotropy of the radiation. Afterwards,
iron needles dominate, degrading the infrared radiation to produce the
observed microwave radiation \cite{Wic06}. Within QSSC, on the other hand, the anisotropies
of this radiation would also be explained in terms of interaction of the radiation with 
clusters of galaxies and other elements \cite{Nar03,Nar07}.

Extragalactic iron whiskers might be formed in a process similar 
to metallic vapours cooling slowly enough in the laboratory. Whiskers are formed by this type
of process during the expansion of the envelopes of supernovae. As a matter of fact,
some of the defenders of the QSSC model \cite{Hoy88} have claimed that iron whiskers are observed
in the emission spectrum of the Crab pulsar PSR0531+21.
Thermalization was also proposed to be due to the plasma of the
intergalactic space, but in this case it takes about 450 Gyr for the starlight
to get thermalized \cite{Ibi06}.

Concerning the origin of the redshift of galaxies, the proposers of this model admit a component due to the
expansion $S(t)$, as in the standard model, but further posit the existence of intrinsic redshifts.  This
allows the solution of problems such as the periodicity of redshift in quasars and the existence of numerous
cases with possible anomalous redshifts \cite{Arp97,Bur01}.

Summing up, QSSC competes with the standard cosmological model to explain many observations,
at least in an approximate way,
but with a very different description of the universe. According to its authors,
QSSC can even explain some facts that the standard model cannot, such as possible anomalies in the redshifts of quasars. It also contains predictions
that are different from those of the standard model, although these are difficult to test. The predictions include
\cite{Nar06}: the existence of faint galaxies ($m>27$) with small blueshifts ($|\Delta z|<0.1$), 
the existence of stars and galaxies older than 14 Gyr, an abundance of baryonic matter
in ratios above those predicted by $\Lambda $CDM, and gravitational radiation derived from the
creation of matter.

\section{Plasma Cosmology}

Plasma Cosmology is another alternative model that has occupied two or three generations of researchers,
some of them still active today. 
Its proponents include the physics Nobel laureate Hannes Alfv\'en, Oskar Klein, 
Anthony L. Peratt, and Eric J. Lerner.
It has a strong argument against one of the main pillars of the standard model: It
proposes an alternative to the belief that gravitation is the fundamental force that controls the dynamics
of the universe. It assumes instead that most of the mass in the universe is plasma controlled mainly by
electromagnetic forces (and also gravity, of course), rather than gravity alone \cite{Alf62,Alf83,Alf81,Alf88,Ler91}. 
According to this theory the universe has always existed, it is always evolving, and it will continue to 
exist forever. 

The plasma in the laboratory, through electric currents and magnetic fields, creates filaments similar to those
observed in the large-scale filamentary structure of the universe. 
The plasma cosmology model predicts the observed morphological hierarchy: distances
among stars, galaxies, cluster of galaxies, and filaments of huge sizes in the large-scale
structure. The observed velocities of the streams of galaxies in regions close to the largest
superclusters are coincident with those predicted by the model, without the need for dark matter \cite{Ler91}. 
The formation of galaxies and their dynamics would also be governed by forces and interactions
of electromagnetic fields \cite{Per83,Per84,Ler91}.

Hubble--Lema\^itre expansion was admitted in the first version of plasma cosmology and was
explained by means of the repulsion between matter and antimatter \cite{Alf62,Alf79}.
A plasma mechanism can separate
matter from antimatter and, when an antimatter cloud bumps into an
ordinary-matter cloud, they will not totally annihilate each other; instead,
only a thin layer  
will be annihilated \cite{Leh77}, generating a hot, low-density plasma
layer which will push the clouds apart.
Alfv\'en proposed his `fireworks' model, in which a supercluster is
repelled by other superclusters; within a supercluster each cluster is repelled by other
clusters; and within a given cluster each galaxy is repelled by the other galaxies, and so on,
obeying a distribution of matter and antimatter.
In each local volume, a small explosion would impose its own local Hubble--Lema\^itre relationship, and
this would explain the variations in the velocities of the Hubble--Lema\^itre law, i.e.\ the different values of the Hubble 
constant measured in the '70s and '80s
when Alfv\'en posited his hypothesis, in different ranges of distances or looking 
in different directions, all without invoking
dark matter. The energy derived from the annihilation of protons and electrons would produce
a background radiation of X- and $\gamma$-rays.
There are some objections against the existence of antimatter  based on
the absence of $\gamma$-rays from anhilation, but they are model-dependent.
Many of the objections against antimatter have been analysed \cite{Rog80}
and it has been shown that none of them is crucial. Nonetheless, some critics remarked that it 
is not consistent with the isotropy of the X-ray backgrounds \cite{Pee93}.

Instead of expansion caused by matter--antimatter repulsion, in more recent times, some proponents
of plasma cosmology \cite{Ler06} have stated that there is no expansion, that the universe is static,
and that the redshift of the galaxies would be explained by some kind of tired-light effect of the
interaction of photons with electrons in the plasma. (See \S \ref{.static})

With regard to the CMBR, Refs. \citen{Ler88,Ler95} 
explain it in terms of absorption and re-emission of radiation produced by stars.
It is similar to the mechanism proposed by  QSSC, but here the
thermalization is due to interaction with electrons. The interaction of photons
and electrons produces a loss of direction in the path of the light, giving rise to
isotropic radiation.

\section{Universe as a Hypersphere}
\label{.hypersp}

Another category of models that have appeared in different versions is one that posits that our universe is
a hypersphere of a higher-dimensionality geometrical entity; that is, a set of points at a constant distance
from its centre, constituting a manifold with one dimension less than that of the ambient space.

One of the first models maintaining this idea is the `Chronometric Cosmology' of Irving E. Segal \cite{Seg76,Seg95}. 
This model assumes that global space structure is a 3D-hypersphere (or 3D-hypersurface, 
as Segal calls it) in a universe of four dimensions. Events in the universe are ordered globally according to a temporal order.
This model makes an application of general relativity different from the standard model,
getting a relationship of the redshift with distance $r$: $z=\tan ^2 \frac{r}{2R}$.
His cosmology gives a good fit to the various curves versus redshift: 
magnitude, counts, angular size, etc., but, with data about redshifts of galaxies and distances, it is now known the proposed 
relationship between redshift and distance cannot be correct \cite{San95}. Moreover,
there is no explanation for the CMBR.
Many other refutations of Segal's claims 
have also been published \cite{Sal86,Wri87,San92,Kor97}.

More recent is the
hypothesis of the existence of five combined spacetime dimensions.
By making some peculiar assignments between coordinates and physical 
distances and time, a hyperspherical symmetry is made apparent by assigning the hypersphere radius to proper time and distances 
on the hypersphere to usual 3-dimensional distances in a Euclidean universe \cite{Alm06}, which can explain the Hubble--Lema\^itre expansion law without  appealing to dark matter; an empty universe will expand naturally at a flat rate in this way. Another variation is the
Hypersphere World-universe Model \cite{Net20,Net21}, which claims the existence of 
a 3-dimensional hypersphere with respect to a 4-dimensional Nucleus of the World. 
Matter in this universe is of the ordinary kind with the addition of a multicomponent dark-matter (instead
of Cold Dark Matter plus Dark Energy). This
model has a number of peculiar characteristics: the beginning of the universe, instead of originating from
a singularity, stems from a 4-dimensional Nucleus of the World; the radius of this Nucleus, increasing with
speed $c$, is what produces the expansion; the CMBR stems from the thermodynamic equilibrium of photons
with intergalactic plasma; and the nucleosynthesis of light elements occurs inside dark matter cores of
Macro-objects. 

\section{Static Models and non-cosmological redshifts}
\label{.static}

At the farthest extreme of alternative models with respect to the standard cosmology, we have proposals
of universe that contradict the main interpretation upon which standard cosmology was created from the 1920s onwards: 
expansion of the universe and the interpretation of redshifts as cosmological. Some versions of plasma and hypersphere cosmologies figure among  these models, but
there exist plenty of other models that are characterized by the lack of an origin of time
(a universe of indefinite age),  no expansion, and in some cases
no limit is imposed on space which is Euclidean.

Of the many cases of static and/or non-cosmological redshift models in the literature, we will mention just
a few. Some of them are totally obsolete and with no researcher actively working on them today, but they
deserve to be mentioned from a historical point of view. Most of them are proposed by single researchers,
they are the author's own theories, in some cases produced by non-active professional individuals within
astrophysics, retired researchers or professionals of other fields; nonetheless, they serve to illustrate
the range of possible speculations within cosmology.

\subsection{Cosmological models motivated by tired-light redshifts}

The redshift of galaxies given by Hubble--Lema\^itre's law can be due to some mechanism different
from the expansion or Doppler effect \cite{Nar89,Bar94}. 
A bibliographical catalogue of non-trivial redshifts collected by Reboul \cite{Reb81} in 1981
reduces hundreds of references into 19 classes
of alternative theories, of which the tired-light scenario is the largest.

A tired-light scenario assumes that the photon loses energy owing to some proposed photon--matter process,
photon--photon interactions, or some dissipative property of the photon. The energy loss can be as a
function of time or distance: the distance can be either long, spanning the intergalactic space
between the object and the earth, or short, for instance taking into consideration only the corona enveloping
the object. The scenario is sometimes presented as a possible phenomenological approach describing photon
energy loss in a putative static universe, sometimes given as a theory for the energy loss mechanism.

There are several hypotheses that can produce this tired-light effect. The idea of loss of energy of the photon in
the intergalactic medium was first suggested in 1929 by Zwicky \cite{Zwi29} and was defended by him for a long time.
As late as the mid-twentieth century,
he maintained that the hypothesis of tired-light was
viable\cite{Zwi57}. Other authors from this epoch also supported the idea \cite{Mac32,Ner37,Fin54}. But there
are two problems \cite{Nar89}: 1)
for a particle to lose energy in an interaction also implies a momentum transfer that
smears out the coherence of the radiation from the source, and so all images of distant objects would look
blurred if intergalactic space produced scattering, something that is incompatible with present-day
observations \cite{Ste07}; 2) the scattering effect and consequent loss of energy would be frequency
dependent, which is again incompatible with what we observe in galaxies \cite{Bar12}. Nonetheless, there are
theories that propose  solutions to mitigate these problems \cite{Vig88,Mar88,Mam10,Roy00}.

\begin{figure}
\vspace{0cm}
\centering
\includegraphics[width=8.7cm]{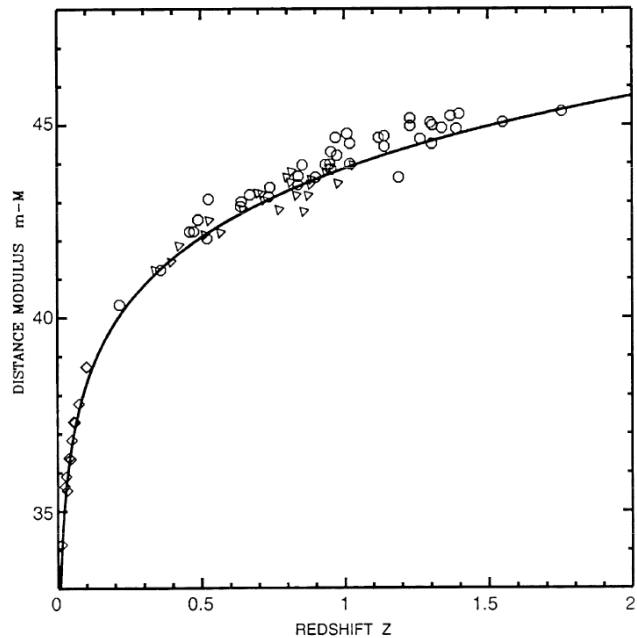}
  \caption{Comparison of the model Hubble diagram for distant Type Ia supernovae in a static tired-light
    cosmology (line) with the Type Ia supernova (SN) data.  The model has no free adjustable parameters. The
    squares represent the low-redshift SN data adopted from Riess et al., the triangles represent IfA Deep
    Survey SN data adopted from Barris et al., and the circles represent the Hubble Space Telescope SN data
    adopted from Riess et al. The high-redshift SN data are subject to uncertain apparent magnitude
    corrections for host galaxy extinction.
    (Adapted from Ref. [\citen{Sor09}] under the Creative Commons Attribution 4.0 International 
License, https://creativecommons.org/licenses/by/4.0/legalcode.)}
    \label{fig:tired-light-Hubble}
\end{figure}

The strong appeal of tired-light redshift is that it could naturally explain many types of non-trivial
redshift phenomenologies such as `redshifts on or by the Sun', `general problems of redshifts', `redshifts
of stars', `morphological redshifts'\cite{Reb81} and `Fingers of God' in the large-scale-structure (usually
explained as gravitational infall velocities that produce Doppler).  By extension contemporary problems such
as the `anisotropy of the Hubble Law',\cite{Col19} `anomalous redshift of quasar-galaxy
associations'\cite{Arp97,Bur01}, `galactic rotation curves', the `Hubble tension' and the `acceleration of
expansion'\cite{Sor09} (see Fig.~\ref{fig:tired-light-Hubble}) can also be easily interpreted as an effect
of a tired-light redshift.

A `static cosmology' common to these models may be described with the Einstein static cosmology,\cite{Mol91}
or with other metrics different from general relativity, although there are also proposals away from
relativistic approaches.
Since in a static universe a cosmological redshift cannot be produced, the tired-light mechanism supplies a differential Hubble law in the form $d\nu/\nu = -(c/H) dr_p$ to get
\begin{equation}
  r_p = (c/H) \ln(1+z),
\end{equation}
where $r_p$ is the proper distance.
From this metric follows these relations: luminosity distance $d_L  = (1+z)^{1/2} \ r_p$,
distance modulus $m(z) = 2.5 \log(1+z) + 5 \log(\ln(1+z)) + 25$,
angular size $\theta \propto \ln(1+z)$,
number counts $\log N(z) = 3 \log[\ln(1+z)] + cte$,
and surface brightness $\log SB(z) = 2.5 \log (1+z)$.

\begin{figure}
\vspace{0cm}
\centering
\includegraphics[width=8.7cm]{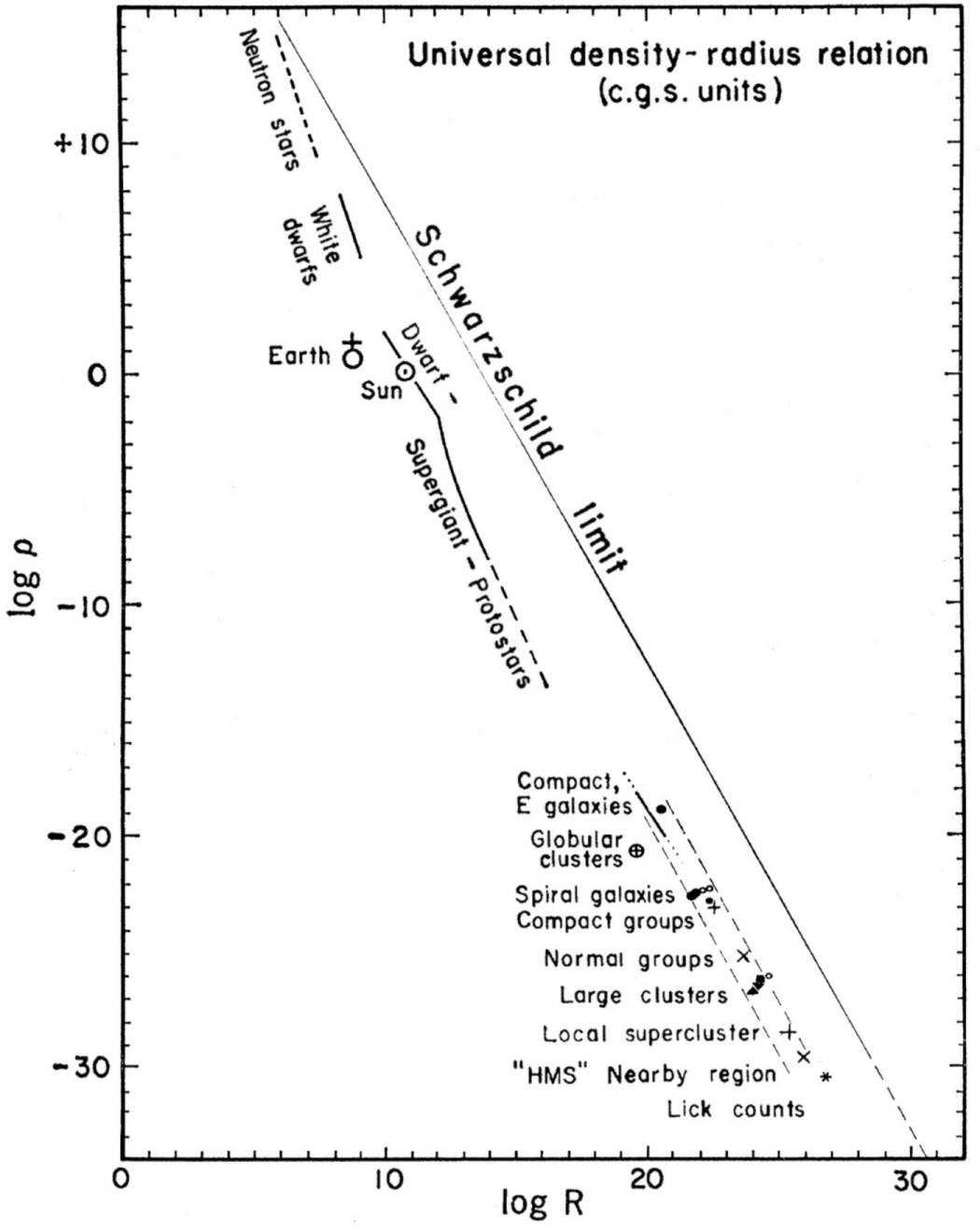}
  \caption{ The hierarchical argument uses the density of the largest structures observed in the universe
    which remain well below the Schwarzschild limit over 40 orders of magnitudes.  The universal
    density-radius relation gives the maximum average density of matter [g/cm$^3$] in spherical volumes of
    radius R [cm] from neutron stars (dashed line at top) to the largest domain in which galaxies have been
    counted (asterisk at bottom). The range of densities by the virial theorem for stellar and galaxy
    clusters is shown (thin dashes).
    (Adapted with permission of the AAAS from Fig.~3 in Ref. [\citen{deV70}].)}
  \label{fig:hierarchy}
\end{figure}

The hierarchical argument\cite{deV70, Jaa75} proposes that the density of the largest
structures observed in the universe remain well below the Schwarzschild limit over 40 orders of
magnitudes (see Fig.~\ref{fig:hierarchy}).  The hierarchical universe gives an easy solution to
the gravitational paradox of Seeliger.  It also implies a curvature radius that is extremely
large $R \gg c/H.$  As a result, proponents of a static universe consider that these observations
forbid the concept where the singularity occurs everywhere at the same time.\cite{Pec97}

The CMBR is interpreted to be that of a thermalization of stellar radiation based on predictions made in the 1950s by Finlay-Freundlich and Max Born on the basis of a tired-light mechanism.\cite{Pec97}
  
  Most of these considerations are used in the following static cosmologies, with some specific variations.

\subsubsection{Curvature Cosmology} 

Proposed by David F. Crawford \cite{Cra11a,Cra11b,Cra11c,Cra18}: 
This cosmology is based on a combination of general relativity and quantum mechanics. 
The concept of curvature pressure arises from the ideas that the density of particles produces curved
spacetime and the velocity of these particles produces a reaction pressure that acts to decrease the local
spacetime curvature.  Although the geometry is similar to the original Einstein static model, this cosmology
differs in that it is stable because of a feedback mechanism: the plasma produces curved space-time through
its density entering the stress-energy tensor in Einstein's field equations, and the velocities of the
plasma particles decrease this curvature. \cite{Cra18}
In the case of the photons that travel across matter, the curvature
produces a tired-light effect as a product of the gravitational interaction
between wave packets and curved spacetime which will follow geodesics and
be subject to geodesic focusing, giving rise to the observed curvature redshift of galaxies.

Since this will alter the transverse properties
of the wave, some of its properties such as angular momentum
would be altered, which is forbidden by quantum mechanics.  Instead,
the result of the interaction of the photon is three new photons: one with almost 
identical energy and momentum to that of the original photon and two extremely low energy secondary photons.
Anomalous redshift cases might be produced by the extra redshift being due to the photons'
passage through the cloud around the anomalous object \cite{Cra11a,Cra11c}. The CMBR comes from the 
curvature--redshift process acting on the high-energy electrons and ions in the cosmic plasma. 
The energy loss which gives rise to the spectrum of photons of the CMBR occurs when an electron that 
has been excited by the passage through curved spacetime interacts with a photon
or charged particle and loses its excitation energy.

\subsubsection{Plasma-Redshift}

Proposed by Ari Brynjolfsson,\cite{Bry06, Bry04} the plasma-redshift is a result of a photon-electron
interaction which follows strictly from the average electron density in intergalactic space.  The
magnitude-redshift relation also includes extinction caused by Compton and Rayleigh scattering on bound
electrons in atoms.  There is no time dilation and no expansion of the universe.

Plasma redshift decreases the photon energy and increases the kinetic energy of the electrons, which
produces the steep temperature rise in the transition zone to million degrees K in the solar corona.
Although the universe is infinite, the gravitational potential has a finite range due to the finite
propagation speed of gravity.  Random thermal motion reduce the ability of the gravitational field to
transfer information about its direction and strength over large distances.
Apart from small ripples (e.g. gravitational lensing) space is flat.

\subsubsection{Subquantum Kinetics}

Proposed by Paul A. LaViolette \cite{LaV12}: This is a unified field theory with the foundations
for a new wave--theory of matter. Its non-dispersing, periodic structures resolve the wave--particle
duality and produce de Broglie wave diffraction effects. The Subquantum kinetics model proposes an open,
order-generating universe, continuously creating matter and energy
in the form of neutrons in the vicinity of existing particles.\cite{LaV12a} 
It predicts that gravitational
potential should have a finite range. It uses tired-light redshift in a static universe, without
radiating a secondary photon, angular deflection, or a strong wavelength dependence. It works as if 
intergalactic space were on average endowed with a negative gravitational mass density.

\subsubsection{Scale Expanding Cosmos} 
Proposed by Carl Johan Masreliez \cite{Mas99,Mas12}: A universe is proposed in which not only space
expands (therefore, it is not properly a static = non-expanding model) but time also expands: the
relationship between space and time could remain constant during the cosmological expansion and all
cosmological locations in time and space could be equivalent, if the metrics of both space and time
expand.  The scale expansion could be eternal, which would eliminate the creation event. Redshift
is a tired-light effect. The theory is presented based on the proposition that all four metric
coefficients of space and time change with cosmological expansion. Such a universal scale expansion
would preserve the four-dimensional spacetime geometry and therefore, by general relativity, most
physical relationships. In addition, if the scale expansion were exponential with time, all epochs
would be equivalent. There are also other proposals on the reinterpretation of space in terms of
time expansion or comoving system of units \cite{Oli02,Oli21}.

\subsubsection{Dichotomous cosmology}
  
Proposed by Yuri Heyman:\cite{Hey14} Contrary to 
general relativity, here there is a dichotomy between
light and matter dynamics: the luminous portion of the Universe is expanding at a constant rate as in the de
Sitter cosmology in a flat Universe, whereas the matter component is static, and the emitted light
wavelength gets stretched due to a tired-light process. As a consequence time-dilation of supernovae light
curves is observed with a stretching factor of $1+z$. In addition, the expanding luminous world is
consistent with the radiation energy density factor $(1+z)^4$ inferred from the CMBR.

\subsubsection{Wave System Cosmology}
Proposed by Thomas B. Andrews \cite{And88,And98}: The universe is a pure system of waves with mass density
and tension parameters proportional to the local intensity of the modes of the waves. The peaks of the
constructive interferences are the elementary particles. 
The energy input to a single proton from all
of the interacting particles in the universe is found to be equal to the mass energy of the proton. Newton's
law of gravitation
derives from the wave modes originating among protons. In this flat model of the universe, the
redshift is produced by a tired-light mechanism.

\subsubsection{Eternal Universe}
Proposed by Gerald S. Hawkins \cite{Haw60,Haw62a,Haw62b}: Based on the
existence of a negative pressure in a cosmic fluid derived from general relativity
(not very different from the role that the cosmological constant has acquired nowadays). 
The main point which differentiates this model from the standard theory is the proposal
that the universe is static, infinite, without an instant of creation, and without expansion.
The redshift of the galaxies is explained as a gravitational effect combined with a slight
amount of intergalactic extinction produced by certain particles located in the space surrounding galaxies,
which is $\sim 10^{-7}$ times the local interstellar absorption per unit
distance. Ref. \citen{Haw93} argues that his model is not unstable, having no tendency to collapse or expand, and that the
CMBR is due to the emission of galactic and intergalactic dust grains. 
Olbers' paradox is solved by means of absorption in clouds of dust.

\subsection{Other non-cosmological redshifts and other static models}

A gravitational redshift in a fractal universe as a source of the redshift of the galaxies \cite{Bar81}, or
theories with the context of alternative views of gravitation can also produce a non-cosmological redshift
without expansion. \cite{Gho97,Bro01,Cra06,Iva06,Kra08,Fis10,Mic16} Instead of new ideas on gravitation, some
researchers explored electromagnetism \cite{Fin54,Ros06,Spa21}.
Furthermore, just to give further examples from the huge literature on the topic, there are explanations of
non-cosmological/non-Doppler redshifts in terms of the local shrinkage of the quantum world \cite{Alf06};
the variation of the speed of light \cite{Sht93}; variation of Planck's constant \cite{Hun10}; time
ac\-cel\-er\-a\-tion or deceleration \cite{Gar01,Tag08}; proposals for a quantum long-time energy redshift
\cite{Urb08} in which the energy is smaller owing to a quantum effect when the photon travels over a very
long time; chronometric cosmology \cite{Seg76,Seg95} (see \S \ref{.hypersp}); the variable mass hypothesis
or secular mass variation of particles \cite{Hoy64,Nar77,Nar93,Bar02b} (see \S \ref{.QSSC} and \S
\ref{.SCC}); etc.
All these proposed mechanisms show us that it is quite possible to construct a cosmological scenario with
non-expansion redshifts. Nonetheless, all these theories are at present just speculation without direct
experimental or observational support.

\subsubsection{Self Creation Cosmology (SCC)}
\label{.SCC}
Proposed by Garth Antony Barber and others \cite{Bar02a,Bar02b,Bar03b,Bar03a},
the self creation cosmology
is an adaptation of the Brans Dicke theory in which 
the conservation requirement is relaxed in order to allow the scalar field to interact 
with matter. SCC can be thought of as general relativity combined with
Mach's principle and
local conservation of energy. In SCC, energy is conserved but energy--momentum is not. 
Particle masses increase with gravitational potential energy and, as a 
consequence, cosmological redshift is caused by a secular, exponential 
increase of particle masses. The universe is 
static and eternal in its Jordan frame and linearly expanding in its Einstein 
frame. Furthermore, as the scalar field adapts the cosmological equations, 
these require the universe to have an overall density of only one third of the critical density while 
remaining spatially flat.
The cosmological redshift is interpreted as a measurement of
the cosmological increase of the atomic masses of the measuring apparatus rather than by a Doppler shift.
The theory also posits a time-slip between 
atomic clock time on the one hand, and gravitational ephemeris and cosmological 
time on the other, which would result in the observed cosmic acceleration.

\subsubsection{Cellular Cosmology}
The cellular cosmology of Conrad Ranzan\cite{Ran14, Ran15} is based on a Dynamic Steady State Universe
model. The universe is an assembly of gravity cells which have a nominal diameter of 
60 Mpc.  The change in
velocity of the aether flow is what produces gravity.  The mechanism that causes the gravitation effect and
sustains the Universe's gravity-cell structure is also the mechanism that causes the wavelength stretching
manifesting as cosmic redshift.  Intrinsic spectral shift occurs with a transit of the photon across/through
any gravity well (sink). It is caused by the difference in propagation velocity between the axial ends of
the photon or wave packet.  The CMBR is the ``now-and-forever'' temperature of the steady state 
universe.

\subsection{Plausibility of static models} 
Static models are usually rejected by most cosmologists.  
However, from a purely theoretical point of view, the representation
of the Cosmos as Euclidean and static is not excluded. Both expanding
and static spaces are possible for the description of the universe. 
Before Einstein and the entry of Riemannian and other non-Euclidean
geometries into physics, there were attempts to describe the
known universe in terms of  Euclidean geometry, but these faced
the problem of justifying a stable equilibrium. Within a relativistic context,
Einstein proposed a static model that included a cosmological 
constant,\cite{Ein17} his `biggest blunder' according to himself.
This model still has problems in guaranteeing stability, but it
might be amenable to some kind of solution.
Ref. \citen{Cra18} solves the stability problem with a feedback mechanism
resulting from the tired-light process itself.
Ref. \citen{Nar93} solves it
within a variation of the Hoyle--Narlikar conformal theory
of gravity, in which small perturbations of the flat Minkowski 
spacetime would lead to small oscillations about the line element 
rather than to a collapse. 
Ref. \citen{Boe07} analyses the stability of the Einstein static 
universe by considering homogeneous scalar perturbations in the context 
of $f(R)$ modified theories of gravity, and it is found 
that a stable Einstein cosmos with a positive cosmological constant 
is possible. Other authors solve the stability problem through
variation of fundamental constants \cite{Van84,Tro87}.
Another idea from Ref.~\citen{Van93} is that hypothetical gravitons 
responsible for gravitational interaction
have a finite cross-sectional area, so that they can only travel a finite 
distance, however great, before colliding with another graviton. 
So the range of the force of gravity would necessarily be limited in this way
and collapse avoided.

Curved geometry (general relativity and its 
modifications) does not conserve the energy--momentum of the gravity field. However, Minkowski space does follow the
conservation of energy--momentum of the gravitational field. One approach with a 
material tensor field in Minkowski space is given in Feynman's gravitation \cite{Fey95}, in which 
space is static but matter and fields can be expanding
in that static space. Also worth mentioning is a model related to modern 
relativistic and quantum field theories of basic fundamental interactions 
(strong, weak, electromagnetic): the relativistic field gravity theory and fractal 
matter distribution in static Minkowski space \cite{Bar08b}.

Olber's paradox for a universe without limits  is an old problem \cite{Bon61}
and also needs subtle
solutions, but extinction, absorption, and re-emission of light, fractal
distribution of density, and
the mechanism which itself produces the redshift of the galaxies might
have something to do with its solution.

\section{Classification according to characteristics}
\label{.clasif}

We have described a representative sample of alternative cosmological models, from the minor variations
with respect to $\Lambda $CDM to the most extreme heterodox proposals of static universes.

Attending to the different features of the models, the following differences can be observed among them:

\begin{description}
\item[Gravity, forces:] general relativity in the standard $\Lambda $CDM and its minor variations or in the cases of inhomogeneous universe, Cold Big Bang, Zero Active Mass and Milne, QSSC and even in some
cases of static universe like Hawkins's Eternal Cosmology, Crawford Curvature Cosmology (together with
quantum mechanics) and Masreliez's Space Expanding Cosmology. However, there are deviations from
general relativity when varying physical constants, alternative gravity or cyclical universes are assumed as major variations.  Plasma Cosmology takes electromagnetic forces as dominant dynamical element, and the plasma-redshift cosmology includes finite propagation speed of gravity. Hypersphere models use different hypersphere geometry.  Dichotomous cosmology states a dichotomy
in the dynamics of matter and light, against general relativity principles. Barber's Self-Creation Cosmology uses Brans Dicke gravitation, and Andrews' Wave System and LaViolette Subquantum's kinematics posit the equivalence of Matter and Waves.  In the Cellular Cosmology, gravity is produced by
the change in velocity of an aether flow.

\item[Expansion:] included in all minor or major variations on the standard model and in QSSC. Plasma Cosmology and Hyperspheres may assume either expansion or non-expansion. The static models include non-expansion, except Masreliez's Scale Expanding Model that assume expansion of space and time.

\item[Age of the universe:] finite (e.g., 13.8 Gyr for $\Lambda $CDM; 14.6 Gyr for Zero Active Mass or Milne cosmologies) in all minor or major variations on the standard model except in the Cyclical Universes that keep a eternal existence. The rest of the models keep also an infinite age of the
universe, except Hyperspheres that keep a finite age and Masreliez's Scale Expanding Model where it can be finite or infinite.

\item[Redshift:] cosmological in all minor or major variations on the standard model. QSSC 
assumes a combination of cosmological and variable mass hypothesis redshifts.  Plasma Cosmology has two options: either Doppler effect due to repulsion among galaxies (due to matter-antimatter interaction), or tired-light hypothesis, a plasma-redshift also included in the static cosmology of that name. Hypersphere  models  have their
own cosmological redshift in term of a geometric interpretation. Hawkins's Eternal Cosmology assumes gravitation redshifts. Barber's Self Creation Cosmology assumes a variable mass hypothesis too. Other static models cited here use a tired-light hypothesis.  In the Cellular Cosmology is related to the
same mechanism that produces gravity, a change of velocity in an aether flow.

\item[Dark elements:] CP violation, inflation (inflaton), dark matter and dark energy are the four
dark nightmares of $\Lambda $CDM. The different variations within standard paradigm may include
some of these elements, with some changes in the minor variations, 
but some cases  do not include them or do not explicitly tell us
about their existence. In the case of Conformal Cyclic Cosmology (one of the major variations
with cyclical universes), there is the extra element of black holes evaporation.
QSSC also keeps dark matter and dark energy, plus another mysterious element: the C-field of matter creation. LaViolette's Subquantum kinetics also contains a continuous creation of matter. Plasma Cosmology gets rid of all these dark forces, provided that strong magnetic
fields (undetected so far) exist. Hypersphere has a multicomponent of dark matter. 
Hawkins' Eternal Universe contains some negative pressure (not very different from the concept of
dark energy nowadays). Other models are not explicit about dark elements existence.

\item[CMBR origin:] decoupling of matter and radiation at $z\approx 1100$ in $\Lambda $CDM or its inhomogeneous
variation. In some major variations like Cold Big Bang or Zero Active Mass, or totally different models like QSSC, Plasma Cosmology or Hypersphere Model, the origin is instead a thermalization of radiation of Population III of stars by particles in the intergalactic medium.
Static models are not clear about it, although some of them point out to dust emission
(Hawkins' Eternal Cosmology), or redshifted high energy
charged particles (Crawford's Curvature Cosmology).  In the Cellular Cosmology, CMBR is the
temperature of a steady-state universe.

\item[Light Element Nucleosynthesis:] primordial in the standard model and its variations, 
except the Cold Big Bang, which admits a combination of primordial and stellar (with stars of population III) nucleosynthesis. For QSSC and Plasma Cosmology, it is pure stellar nucleosynthesis. In the version of hypersphere universe by Netchitailo\cite{Net20,Ner21}, it is produced in the dark cores of Macro-objects.

\item[Homogeneity at large-scale:] in all cases except in the major variation
of explicitly inhomogeneous universe, or in some cases with alternative gravity scenarios.

\item[Galaxy formation:] by evolution of density fluctuations with gravity as the only force when
explicitly established. This evolution might be quicker in scenarios with alternative gravity. The exception is Plasma Cosmology, where electromagnetic 
forces govern the dynamics.

\end{description}

\section{General problems of the alternative models}
\label{.caveats}

None of the alternative cosmological models is
as competitive as the standard $\Lambda $CDM model, because they are not so developed and 
there are many observations pending to be explained with these models.
This does not mean that $\Lambda $CDM is necessarily the correct model of the universe, it also
has a bunch of problems pending to be solved \cite{Lop17c,Lop22,Per21}, and many dark elements from which we know nothing. 
However, the level of accuracy of the representation
of the set of cosmological data is much higher and none of the caveats is conclusive so far as to
falsify this model. Moreover, the alternative proposals have some problems too, and they are yet more severe
than in the standard model (see, for instance, Edward L. Wright's
web-page\footnote{http://www.astro.ucla.edu/$\sim $wright/errors.html}), perhaps because these theories are
not as developed and polished as the standard model.

A solution to Olbers' paradox by dust absorption in a universe of no defined extent, for
instance, is not clear. One may wonder, if energy does not disappear, whether the absorbing element (dust)
should be heated and re-emit, and, if the energy disappears how that can be consistent with known physical
laws. This problem has no easy solution.

Expansion in universes without a time origin is either
taken as a fact, in which case the models need speculative elements to argue that there was no 
beginning of the universe, or an alternative explanation
must be given  for the redshift of the galaxies  which raises its own set of difficulties.

Light element abundances require  in some cases  very  early stellar
populations (Population III) that have not been observed yet.
In an indefinitely old universe however, there is more than enough time to form all heavy elements and Pop. III stars does not need to be hypothesized.  In that case their non-observation is not a problem.

The CMBR has alternative explanations to $\Lambda $CDM's,
but with ad hoc elements without direct proof, such as hypothetical particles to thermalize 
stellar radiation. 
A power spectrum with oscillations is a rather 
normal characteristic expected from any fluid with clouds of overdensities that emit/absorb radiation or interact gravitationally with photons, and with a finite range of sizes and distances for those clouds \cite{Lop13b}. 
The standard cosmological interpretation of `acoustic' peaks, from the hypothesis of 
primaeval adiabatic perturbations in an expanding universe, is just a particular 
case; peaks in the power spectrum might be generated in 
scenarios that have nothing to do with oscillations owing to gravitational compression 
in a fluid. Nonetheless,
all proposals to explain a CMBR produced 
in the intergalactic medium---even assuming that a perfect black body shape can be produced---have the problem that the integration 
along the line of sight gives a superposition of many layers of black body radiations, each with a different
redshift, giving in total something different from a black body.
Otherwise the CMBR can originate from the local universe ($z\approx 0$; in such a case,   
the problem would be that space would be too opaque to allow the observations
of distant radio sources \cite{Wri95} and it could not explain the Sunyaev-Zel'dovich effect
of the interaction with clusters of galaxies \cite{Lop17}) or at a given high redshift $z>0$ but 
within a layer with small $\Delta z$.

The most elaborate alternative models, such as QSSC, do indeed apply the same methodology as the standard
model:
each one has some basic tenets, a lot of free parameters and ad hoc elements that are introduced every
time some observation does not fit the model. The modern version of QSSC,
for example, is able to explain most of the difficulties of the previous (Steady State) version of the model.
The authors introduced ad hoc elements without observational support in the same way that the standard
model introduces ad hoc non-baryonic dark matter, dark energy, inflation, etc. The very idea of continuous
creation of matter\footnote{The idea of continuous creation of matter as a result of a modification of
Einstein's equation of gravitational field is also independently explored by other alternative cosmological
models \cite{Yan16}.} also necessitates some very exotic physics, and
has no empirical support. But the authors continued ad hoc to skip over the inconsistencies;
for instance, the maximum redshift of a galaxy was set to be 5 in the initial version of QSSC; however the model's free parameters were conveniently changed when some new observation does not fit the
initial predictions.  In the end, then, the authors can introduce ad hoc corrections that render
their theory compatible with any maximum redshift of a galaxy.

One should not, however, judge any theory according the number of observations that it can successfully
explain, but by the plausibility of its principles and its potential to fit data 
(provided that we have an army of theoreticians able to correct
the theory ad hoc every time new observations need to be accommodated). A pluralist approach to cosmology \cite{Lop22} 
is a reasonable option when the preferred theory is still under discussion. Therefore, given the number of problems
with the standard model, it is quite reasonable to keep a weather eye on alternative
ideas that might at least provide better partial explanations
or interpretations  of certain observed phenomena. Nonetheless, a global 
cosmological theory that
provides a satisfactory explanation of astronomical observations  does not yet exist according to either
standard or in alternative viewpoints.


\section*{Acknowledgments}


\begin{thebibliography}{100}

\bibitem{Lop17c}
M.~L\'opez-Corredoira, {\em Foundations of Physics} {\bf 47}  (2017) 711.

\bibitem{Lop22}
M.~L\'opez-Corredoira, {\em Fundamental Ideas in Cosmology. Scientific,
  philosophical and sociological critical perspectives} (IOP Science, London,
  to be published in 2022).

\bibitem{Per21}
L.~Perivolaropoulos and F.~Skara, {\em arXiv.org} (2021) 2105.05208.

\bibitem{Tau97}
G.~Taubes, {\em Science} {\bf 278}  (1997)   226.

\bibitem{Geo74}
H.~{Georgi}, H.~R. {Quinn} and S.~{Weinberg}, {\em Phys. Rev. Lett.} {\bf 33}
  (1974) 451.

\bibitem{Lop92}
J.~L. {L\'opez}, D.~V. {Nanopoulos} and H.~{Pois}, { {Proton decay and
  cosmology strongly constrain the minimal SU(5) supergravity model}}, in {\em
  Proceedings of the XXVI International Conference on High Energy Physics. Vol.
  II\/}, American Institute of Physics Conference Series Vol.~272 (AIP,
  1992), p. 1395.

\bibitem{Tan20}
M.~{Tanaka}, K.~{Abe}, C.~{Bronner} {\em et~al.}, {\em Phys. Rev. D} {\bf 101}
  (2020) 052011.

\bibitem{Bru17}
S.~Bruggisser, T.~Konstandin and G.~Servant, {\em J. Cosmol. Astropart. Phys.}
  {\bf 2017(11)}  (2017)   id. 34.

\bibitem{Ber19}
F.~U. Bernlochner, C.~Englert, C.~Hays, K.~Lohwasser, H.~Mildner,
  A.~Pilkington, D.~D. Price and M.~Spannowsky, {\em Phys. Lett. B} {\bf 790}
  (2019) 372.

\bibitem{T2K20}
c.~T2K~Collaboration, {\em Nature} {\bf 580}  (2020) 339.

\bibitem{Lid99}
A.~R. Liddle, An introduction to cosmological inflation, in {\em High Energy
  Physics and Cosmology\/},  eds. A.~Masiero, G.~Senjanovic and A.~Smirnov
  (World Scientific, Singapore, 1999) p. 260.

\bibitem{Pee89}
P.~J.~E. Peebles, Inflation and the baryon isocurvature model, in {\em Large
  Scale Structure and Motions in the Universe\/},  eds. M.~Mezzetti,
  G.~Giuricin, F.~Mardirossian and M.~Ramella (Springer, Dordrecht, 1989) p.
  119.

\bibitem{Ste11}
P.~J. Steinhardt, {\em Scientific American} {\bf 304}  (2011) 18.

\bibitem{Cho19}
D.~Chowdhury, J.~Martin, C.~Ringeval and V.~Vennin, {\em Phys. Rev. D} {\bf
  100}  (2019) id. 083537.

\bibitem{Mart13}
J.~Martin, C.~Ringeval and V.~Vennin, {\em Physics of the Dark Universe} {\bf
  5}  (2014) 75.

\bibitem{Ilj17}
A.~Iljas, P.~J. Steinhardt and A.~Loeb, {\em Scientific American} {\bf 316}
  (2017) 32.

\bibitem{Gut17}
A.~H. Guth, K.~Kaiser, A.~D. Linde {\em et~al.}, {\em Scientific American} {\bf
  317}  (2017) 5.

\bibitem{Mel18}
F. Melia and M. L\'opez-Corredoira, {\em Astron. Astrophys.} {\bf
  610}  (2018) A87.

\bibitem{Pat08}
P. Patrick, N. Pinto-Neto, {\em Phys. Rev. D} {\bf 78} (2008) id. 063506.

\bibitem{Bra08}
R. Brandenberger, {\em Physics Today} {\bf 61} (2008) 44.

\bibitem{Nie16}
J. T.~{Nielsen}, A.~{Guffanti}, S.~{Sarkar}, {\em Sci Rep.} {\bf 6} (2016) 35596.

\bibitem{Col19}
J.~{Colin}, R.~{Mohayaee}, M.~{Rameez}, S.~{Sarkar}, {\em Astron. Astrophys.} {\bf 631} (2019) L13.

\bibitem{Mil15}
P. A. Milne, R. J. Foley, P. J. Brown and G. Narayan,
  {\em Astrophys. J.}, {\bf 803} (2015), id. 20.

\bibitem{Kan19}
Y. Kang, Y.-W. Lee, Y.-L. Kim, C. Chung, C. and C. H. Ree,
  {\em Astrophys. J.}, {\bf 889} (2020), id. 8.

\bibitem{Lee20}
Y.-W. Lee, C. Chung, Y. Kang and M. J. Lee,
  {\em Astrophys. J.}, {\bf 903} (2020), id. 22.

\bibitem{Cal98}
R. R.~{Caldwell}, R.~{Dave}, P. J.~{Steinhardt}, {\em Phys. Rev. Lett.} {\bf 80} (1998), 1582.

\bibitem{Tri17}
A. Tripathi, A. Sangwan and H. K. Jassal, 
  {\em J. Cosmol. Astropart. Phys.} {\bf 6} (2017), id. 012.

\bibitem{Ten21}
Y.-P. Teng, W. Lee and K.-W. Ng,
  {\em Phys. Rev. D} {\bf 104} (2021), 083519.

\bibitem{Too81}
A.~Toomre, What amplifies the spirals, in {\em The Structure and Evolution of
  Normal Galaxies\/},  eds. D.~M. Fall and D.~Lynden-Bell (Cambridge University
  Press, Cambridge (U.K.), 1981) p. 111.

\bibitem{Lop02b}
M.~L\'opez-Corredoira, J.~Betancort-Rijo and J.~E. Beckman, {\em Astron.
  Astrophys.} {\bf 386}  (2002) 169.

\bibitem{Lop99}
M.~L\'opez-Corredoira, J.~E. Beckman and E.~Casuso, {\em Astron. Astrophys.}
  {\bf 351}  (1999) 920.

\bibitem{Ben17}
D.~Benhaiem, M.~Joyce and F.~Sylos~Labini, {\em Astrophys. J.} {\bf 851}
  (2017) id. 19, 10 pp.

\bibitem{Bat00}
E.~Battaner and E.~Florido, {\em Fund. Cosmic Phys.} {\bf 21}  (2000) 1.

\bibitem{Fen15}
J.~Q. Feng and C.~F. Gallo, {\em Phys. Int} {\bf 6}  (2015) 11.

\bibitem{Sip21}
A.~Sipols and A.~Pavlovich, {\em Galaxies} {\bf 8}  (2021) id. 36.

\bibitem{Dre05}
J.~Drexler, {\em arXiv.org}   astro-ph/0504512 (2005).

\bibitem{May12}
F.~J. Mayer and J.~R. Reitz, {\em Int. J. Theor. Phys.} {\bf 51}  (2012) 322.

\bibitem{Haj14}
D.~S. Hajdukovic, {\em Physics of the Dark Universe} {\bf 3}  (2014) 34.

\bibitem{Coo05}
F.~I. Cooperstock and S.~Tieu, {\em arXiv.org} astro-ph/0507619 (2005).

\bibitem{Hat95}
N. Hata, R. J. Scherrer, G. Steigman, D. Thomas, T. P. Walker, S. Bludman and P. Langacker,
{\em Phys. Rev. Lett.}, {\bf 75} (1995), 3977.

\bibitem{And10}
M. E. Anderson and J. N. Bregman, J. N.,
{\em Astrophys. J.}, {\bf 714} (2010), 320. 

\bibitem{McG10}
S. S. McGaugh, J. M. Schombert, W. J. G. de Blok and M. J. Zagursky, M. J.,
{\em Astrophys. J. Lett.}, {\bf 708} (2010), L14.

\bibitem{Oli91}
K. A. Olive, {\em Science}, {\bf 251} (1991), 1194.

\bibitem{Bon16}
S. A. Bonometto and R. Maisini,
{\em Universe}, {\bf 4} (2016), 32.

\bibitem{Guo11}
Q.~Guo, S.~White, M.~Boylan-Kolchin {\em et~al.}, {\em Mon. Not. R. Astron.
  Soc.} {\bf 413}  (2011) 101.

\bibitem{Bod01} 
P. Bode, J. P. Ostriker and N. Turok, 
{\em Astrophys. J.} {\bf 556} (2001) 93.

\bibitem{Bos19}
S.~Bose, M.~Vogelsberger, J.~Zavala, C.~Pfrommer, F.-Y. Cyr-Racine, S.~Bohr and
  T.~Bringmann, {\em Mon. Not. R. Astron. Soc.} {\bf 487}  (2019) 522.

\bibitem{Man15}
A.~Manrique and E.~Salvador-Sol\'e, {\em Astrophys. J.} {\bf 803}  (2015) id.
  103.

\bibitem{Whi78}
S.~D.~M. White and M.~J. Rees, {\em Mon. Not. R. Astron. Soc.} {\bf 183}
  (1978) 341.

\bibitem{Bar94}
Y.~V. Baryshev, F.~Sylos~Labini, M.~Montuori and L.~Pietronero, {\em Vistas in
  Astronomy} {\bf 38}  (1994) 419.

\bibitem{Gab05}
A.~Gabrielli, F. Sylos~Labini, M.~Joyce and L.~Pietronero, {\em Statistical Physics
  for Cosmic Structures} (Springer Verlag, Berlin, 2005).

\bibitem{Wil09}
D.~L. Wiltshire, {\em Phys. Rev. D} {\bf 80}  (2009)   id. 123512.

\bibitem{Rom07}
A.~E. Romano, {\em Phys. Rev. D} {\bf 75}  (2007)   id. 043509.

\bibitem{Lay90}
D.~Layzer, {\em Cosmogenesis} (Oxford University Press, Oxford, 1990).

\bibitem{Agu99}
A.~N. Aguirre, {\em Astrophys. J.} {\bf 521}  (1999) 17.

\bibitem{Agu00}
A.~N. Aguirre, {\em Astrophys. J.} {\bf 533}  (2000) 1.

\bibitem{Gol92}
T.~Goldman, J.~P\'erez-Mercader, F.~Cooper and M.~M. Nieto, {\em Phys. Lett B}
  {\bf 281}  (1992) 219.

\bibitem{Dir74}
P.~A.~M.~Dirac, {\em Proc. R. Soc. Lond. A.} {\bf 338} (1974) 439.

\bibitem{Alb99}
A.~Albrecht and J.~Magueijo, {\em Phys. Rev. D} {\bf 59}  (1999) id. 043516.

\bibitem{Sho04}
H.~Shojaie and M.~Farhoudi, {\em Canadian J. Phys.} {\bf 84(10)}  (2004) 933.

\bibitem{Ell07}
G.~R. Ellis, {\em General Relativity and Gravitation} {\bf 39}  (2007) 511.

\bibitem{Unz15}
A.~Unzicker, {\em Einstein's Lost Key: How We Overlooked the Best Idea of the
  20th Century} (CreateSpace Independent Publishing Platform, 2015).

\bibitem{Gup20}
R.~P. Gupta, {\em Mon. Not. R. Astron. Soc.} {\bf 498}  (2020) 4481.

\bibitem{Goh17}
H.~Gohar, {\em Universe} {\bf 3}  (2017)  ~26.

\bibitem{Cli12}
T.~Clifton, P.~G. Ferreira, A.~Padilla and C.~Skordis, {\em Phys. Reports} {\bf
  513}  (2012) 1.

\bibitem{San02}
R.~H. Sanders and S.~S. McGaugh, {\em Annu. Rev. Astron. Astrophys.} {\bf 40}
  (2002) 263.

\bibitem{Fam12}
B.~Famaey and S.~McGaugh, {\em Living Rev. Relativity} {\bf 15}  (2012) id. 10.

\bibitem{San15}
R.~H. Sanders, {\em Canadian J. Phys.} {\bf 93}  (2015) 126.

\bibitem{Mer20}
D.~Merritt, {\em A philosophical approach to MOND. Assessing the Milgromian
  Research Program in Cosmology} (Cambridge Univ. Press, Cambridge, 2020).

\bibitem{Lel17}
F.~Lelli, S.~S. McGaugh, J.~M. Schombert and M.~S. Pawlowski, {\em Astrophys.
  J.} {\bf 836}  (2017) id. 152.

\bibitem{Bek84}
J.~Bekenstein and M.~Milgrom, {\em Astrophys. J.} {\bf 286}  (1984) 7.

\bibitem{Mil10}
M.~Milgrom, {\em Mon. Not. R. Astron. Soc.} {\bf 403}  (2010) 886.

\bibitem{Bek04}
J.~D. Bekenstein, {\em Phys. Rev. D} {\bf 70}  (2004)   id. 083509.

\bibitem{Fel84}
J.~E. Felten, {\em Astrophys. J.} {\bf 286}  (1984) 3.

\bibitem{San98}
R.~H. Sanders, {\em Mon. Not. R. Astron. Soc.} {\bf 296}  (1998) 1009.

\bibitem{San06}
R.~H. Sanders, Mond and cosmology, in {\em Mass Profiles and Shapes of
  Cosmological Structures (EAS Publ. Ser. 20)\/},  eds. G.~A. Mamon, F.~Combes,
  C.~Deffayet and B.~Fort (EDP Sciences, Les Ulis Cedex, 2006) p. 231.

\bibitem{Sko21}
C.~{Skordis}, T.~{Z\l o\'snik}, {\em Phys. Rev. Lett.} {\bf 127} (2021) id. 161302.

\bibitem{Her12}
X.~Hern\'andez, M.~A. Jim\'enez and C.~Allen, {\em Eur. Phys. J. C} {\bf 72}
  (2012)   id. 1884.

\bibitem{Pov04}
A.~Poveda and C.~Allen, {\em Rev. Mex. Astron. Astrophys. Conf. Ser.} {\bf 21}
  (2004) 49.

\bibitem{Jia10b}
Y.-F. Jiang and S.~Tremaine, {\em Mon. Not. R. Astron. Soc.} {\bf 401}  (2010)
  977.

\bibitem{Mof06}
J.~W. Moffat, {\em J. Cosmol. Astropart. Phys.} {\bf 2006(3)}  (2006)   id.
  004.

\bibitem{Sot10}
T.~P. Sotiriou and V.~Faraoni, {\em Rev. Mod. Phys.} {\bf 82}  (2010) 451.

\bibitem{Kra08}
K.~Krasnov and Y.~Shtanov, {\em Class. and Quantum Gravity} {\bf 25}  (2008)
  id. 5002.

\bibitem{Ann16}
A.~Annila, {\em Entropy} {\bf 18}  (2016)   191.

\bibitem{Jac14}
L.~S. Jackson, {\em Weyl Conformal Gravity: Tests and New Solutions} (Lambert
  Academic Publishing, Chisinau, 2014).

\bibitem{Gor88}
T.~G\"ornitz, {\em Int. J. Theor. Phys.} {\bf 27}  (1988) 659.

\bibitem{Joh96}
M.~V. John and K.~B. Joseph, {\em Phys. Lett. B} {\bf 387}  (1996) 466.

\bibitem{Mel12}
F.~Melia and A.~S. Shevchuk, {\em Mon. Not. R. Astron. Soc.} {\bf 419}  (2012)
  2579.

\bibitem{Mel15}
F.~Melia and M.~L\'opez-Corredoira, {\em Int. J. Mod. Phys. D} {\bf 26}  (2017)
id. 1750055.

\bibitem{Mel17}
F.~Melia, {\em Frontiers of Phys.} {\bf 12} (2017) 129802.

\bibitem{Mel19}
F.~Melia, {\em Mon. Not. R. Astron. Soc.} {\bf 489} (2019) 517.

\bibitem{Pla20}
Planck~Collaboration, {\em Astron. Astrophys.} {\bf 641}  (2020) id. A6.

\bibitem{Mel20}
F.~Melia, {\em Eur. Phys. J. Plus} {\bf 135}  (2020) id. 511.

\bibitem{Kim16}
J.~Kim, P.~Naselsky and M.~Hansen, {\em Mon. Not. R. Astron. Soc.} {\bf 460}
  (2016) L119.

\bibitem{Ben08}
A.~Benoit-L\'evy and G.~Chardin, {\em arXiv} {\bf .}  (2008)   0811.2149.

\bibitem{Cha15}
O.~I. Chashchina and Z.~K. Silagadze, {\em Universe} {\bf 1}  (2015) 307.

\bibitem{Vis13}
R.~G. Vishwakarma, {\em Phys. Scr.} {\bf 87}  (2013)   id. 055901.

\bibitem{Pen10}
R.~Penrose, {\em Cycles of Time: An Extraordinary New View of the Universe}
  (Bodley Head, London, 2010).

\bibitem{Ste07b}
P.~J. Steinhardt and N.~Turok, {\em Endless Universe. Beyond the Big Bang}
  (Orion, London, 2008).

\bibitem{Sun20}
T.~Suntola, {\em J. Phys. Conf. Ser.} {\bf 1466}  (2020) id. 012003.

\bibitem{Hoy48}
F.~Hoyle, {\em Mon. Not. R. Astron. Soc.} {\bf 108}  (1948) 372.

\bibitem{Bon48}
H.~Bondi and J.~Gold, {\em Monthly Notices of the Royal Astronomical Society}
  {\bf 108}  (1948) 252.

\bibitem{ORa14}
C.~O'Raifeartaigh, B.~McCann, W.~Nahm and S.~Mitton, {\em European Phys. J. H}
  {\bf 39}  (2014) 353.

\bibitem{Bur57}
E.~M. Burbidge, G.~R. Burbidge, W.~A. Fowler and F.~Hoyle, {\em Reviews of
  Modern Physics} {\bf 29}  (1957) 547.

\bibitem{Ryl61}
M.~Ryle and R.~W. Clarke, {\em Monthly Notices of the Royal Astronomical
  Society} {\bf 122}  (1961) 349.

\bibitem{Hoy93}
F.~Hoyle, G.~Burbidge and J.~V. Narlikar, {\em Astrophys. J.} {\bf 410}  (1993)
  437.

\bibitem{Hoy94}
F.~Hoyle, G.~Burbidge and J.~V. Narlikar, {\em Mon. Not. R. Astron. Soc.} {\bf
  267}  (1994) 1007.

\bibitem{Hoy00}
F.~Hoyle, G.~R. Burbidge and J.~V. Narlikar, {\em A different approach to
  Cosmology} (Cambridge Univ. Press, Cambridge, 2000).

\bibitem{Nar07}
J.~V. Narlikar, G.~Burbidge and R.~G. Vishwakarma, {\em J. Astrophys. Astron.}
  {\bf 28}  (2007) 67.

\bibitem{Nay99}
A.~Nayeri, S.~Engineer, J.~V. Narlikar and F.~Hoyle, {\em Astrophys. J.} {\bf
  525}  (1999) 10.

\bibitem{Wri95}
E.~L. Wright, {\em Mon. Not. R. Astron. Soc.} {\bf 276}  (1995) 1421.

\bibitem{Hoy95}
F.~Hoyle, G.~Burbidge and J.~V. Narlikar, {\em Mon. Not. R. Astron. Soc.} {\bf
  277}  (1995) L1.

\bibitem{Wic06}
N.~C. Wickramasinghe, Evidence for iron whiskers in the universe, in {\em
  Current issues in Cosmology\/}, eds. J.-C. Pecker and J.~V. Narlikar
  (Cambridge University Press, Cambridge (U. K.), 2006) p. 152.

\bibitem{Nar03}
J.~V. Narlikar, R.~G. Vishwakarma, A.~Hajian, T.~Souradeep, G.~Burbidge and
  F.~Hoyle, {\em Astrophys. J.} {\bf 585}  (2003) 1.

\bibitem{Hoy88}
F.~Hoyle and N.~C. Wickramashinghe, {\em Astrophys. Space Sci.} {\bf 147}
  (1988) 245.

\bibitem{Ibi06}
M.~Ibison, Thermalization of starlight in the steady-state cosmology, in {\em
  1st Crisis in Cosmology Conference (AIP Conf. Ser. 822(1))\/},  eds. E.~J.
  Lerner and J.~B. Almeida (AIP, Melville, 2006) p. 171.

\bibitem{Arp97}
H.~Arp, {\em J. Astrophys. Astron.} {\bf 18(4)} (1997) 393.

\bibitem{Bur01}
G. R. Burbidge, {\em Publ. Astron. Soc. Pacific} {\bf 113} (2001) 899.

\bibitem{Nar06}
J.~V. Narlikar, The quasi-steady-state cosmology, in {\em Current issues in
  Cosmology\/},  eds. J.~C. Pecker and J.~V. Narlikar (Cambridge University
  Press, Cambridge (U. K.), 2006) p. 139.

\bibitem{Alf62}
H.~Alfv\'en and O.~Klein, {\em Arkiv f\"or fysik} {\bf 23}  (1962)   187.

\bibitem{Alf83}
H.~Alfv\'en, {\em Astrophys. Space Sci.} {\bf 89}  (1983) 313.

\bibitem{Alf81}
H.~Alfv\'en, {\em Cosmic Plasma (Astrophys. Space Sci. Library, vol.
  82)} (Reidel, Dordrecht, 1981).

\bibitem{Alf88}
H.~Alfv\'en, {\em Laser and Particle Beams} {\bf 16}  (1988) 389.

\bibitem{Ler91}
E.~J. Lerner, {\em The Big Bang never happened: a startling refutation of the
  dominant theory of the origin of the universe} (Random House, Toronto, 1991).

\bibitem{Per83}
A.~L. Peratt, {\em Sky and Telescope} {\bf 66}  (1983) 19.

\bibitem{Per84}
A.~L. Peratt, {\em Sky and Telescope} {\bf 68}  (1984) 118.

\bibitem{Alf79}
H.~Alfv\'en, {\em Astrophys. Space Sci.} {\bf 66}  (1979) 23.

\bibitem{Leh77}
B.~Lehnert, {\em Astrophys. Space Sci.} {\bf 46}  (1977) 61.

\bibitem{Rog80}
S.~Rogers and W.~B. Thompson, {\em Astrophys. Space Sci.} {\bf 71}  (1980) 257.

\bibitem{Pee93}
P.~J.~E. Peebles, {\em Principles of Physical Cosmology} (Princeton Univ.
  Press, Princeton, 1993).

\bibitem{Ler06}
E.~J. Lerner, Evidence for a non-expanding universe: Surface brightness data
  from hudf, in {\em 1st Crisis in Cosmology Conference\/},  eds. E.~J. Lerner
  and J.~B. Almeida (AIP Conf. Ser. 822(1)). AIP, Melville (2006) p.
  60.

\bibitem{Ler88}
E.~J. Lerner, {\em Laser and Particle Beams} {\bf 6}  (1988) 457.

\bibitem{Ler95}
E.~J. Lerner, {\em Astrophys. Space Sci.} {\bf 227}  (1995) 61.

\bibitem{Seg76}
I.~E. Segal, {\em Mathematical Cosmology and Extragalactic Astronomy} (Academic
  Press, New York, 1976).

\bibitem{Seg95}
I.~E. Segal and Z.~Zhou, {\em Astrophys. J. Supp. Ser.} {\bf 100}  (1995) 307.

\bibitem{San95}
A.~Sandage and G.~A. Tammann, {\em Astrophys. J.} {\bf 446}  (1995) 1.

\bibitem{Sal86}
E.~E. Salpeter and G.~L. Hoffman, {\em Proc. US Nat. Acad. Sci.} {\bf 83}
  (1986) 3056.

\bibitem{Wri87}
E.~L. Wright, {\em Astrophys. J.} {\bf 313}  (1987) 551.

\bibitem{San92}
A.~Sandage, {\em Physica Scripta} {\bf T43}  (1992) 22.

\bibitem{Kor97}
D.~M. Koranyi and M.~A. Strauss, {\em Astrophys. J.} {\bf 477}  (1997) 36.

\bibitem{Alm06}
J.~Almeida, Geometric drive of the universe's expansion, in {\em 1st Crisis in
  Cosmology Conference (AIP Conf. Ser. 822(1))\/},  eds. E.~J. Lerner and J.~B.
  Almeida (AIP, Melville, 2006) p. 110.

\bibitem{Net20}
V.~S. Netchitailo, {\em J. High Energy Phys., Gravitation and Cosmology} {\bf
  6}  (2020) 133.

\bibitem{Net21}
V.~S. Netchitailo, {\em J. High Energy Phys., Gravitation and Cosmology} {\bf
  7}  (2021) 915.

\bibitem{Nar89}
J.~V. Narlikar, {\em Space Science Reviews} {\bf 50}  (1989) 523.

\bibitem{Reb81}
H.~J. Reboul, {\em Astron. Astrophys. Supp. Ser.} {\bf 45}  (1981) 129.

\bibitem{Zwi29}
F.~Zwicky, {\em Proc. Nat. Acad. Sci. USA} {\bf 15}  (1929) 773.

\bibitem{Zwi57}
F.~Zwicky, {\em Morphological Astronomy} (Springer, Berlin, 1957).

\bibitem{Mac32}
W.~D. MacMillan, {\em Nature} {\bf 129}  (1932)  93.

\bibitem{Ner37}
W.~von Nernst, {\em Zeit. Phys} {\bf 106}  (1937) 633.

\bibitem{Fin54}
E.~Finlay-Freundlich, {\em Proc. Phys. Soc. A} {\bf 67}  (1954) 192.

\bibitem{Ste07}
E.~Steinbring, {\em Astrophys. J.} {\bf 655}  (2007) 714.

\bibitem{Bar12}
Y.~V. Baryshev and P.~Teerikorpi, {\em Fundamental Questions of Practical
  Cosmology} (Springer Verlag, Dordrecht, 2012).

\bibitem{Vig88}
J.~P. Vigier, Alternative interpretation of the cosmological redshift in terms
  of vacuum gravitational drag, in {\em New Ideas in Astronomy\/},  eds.
  F.~Bertola, B.~Madore and J.~Sulentic (Cambridge University Press, Cambridge,
  1988) p. 257.

\bibitem{Mar88}
P.~Marmet, {\em Phys. Essays} {\bf 1}  (1988) 24.

\bibitem{Mam10}
D.~L. Mamas, {\em Phys. Essays} {\bf 23}  (2010) 326.

\bibitem{Roy00}
S.~Roy, M.~Kafatos and S.~Datta, {\em Astron. Astrophys.} {\bf 353}  (2000)
  1134.

\bibitem{Sor09}
W.~H.~Sorrell, {\em Astrophys. Space Sci.} {\bf 323(2)} (2009) 205.

\bibitem{Mol91}
M.~Moles, {\em The Physical Universe: The Interface Between Cosmology, Astrophysics and Particle Physics}
  (Berlin, Heidelberg, 1991) 197.

\bibitem{deV70}
G.~de~Vaucouleurs, {\em Science} {\bf 167(3922)} (1970) 1203.

\bibitem{Jaa75}
T.~Jaakkola, M.~Moles, J.~P.~Vigier, J.C.~Pecker and W.~Yourgrau, {\em Found. Phys.} {\bf 5(2)} (1975) 257.

\bibitem{Pec97}
J.-C.~Pecker, {\em Journal of Astrophysics and Astronomy} {\bf 18(4)} (1997) 323.

\bibitem{Cra11a}
D.~F. Crawford, {\em J. Cosmology} {\bf 13}  (2011) 3875.

\bibitem{Cra11b}
D.~F. Crawford, {\em J. Cosmology} {\bf 13}  (2011) 3947.

\bibitem{Cra11c}
D.~F. Crawford, {\em J. Cosmology} {\bf 13}  (2011) 4000.

\bibitem{Cra18}
D.~F.~Crawford, {\em arXiv.org}  1804.10274 (2019).

\bibitem{Bry06}
A.~Brynjolfsson, {\em arXiv.org} astro-ph/0605599 (2006).

\bibitem{Bry04}
A.~Brynjolfsson, {\em arXiv.org} astro-ph/0411666 (2004).

\bibitem{LaV12}
P.~A. LaViolette, {\em Subquantum kinetics: The Alchemy of Creation}, 4th edn.
(Starlane Publ, Niskayana (NY), US, 2012).

\bibitem{LaV12a}
P.~A.~LaViolette, {\em Physics Procedia} {\bf 38} (2012) 326.

\bibitem{Mas99}
C.~J. Masreliez, {\em Astrophys. Space Sci.} {\bf 266}  (1999) 399.

\bibitem{Mas12}
C.~J. Masreliez, {\em The Progression of Time} (CreateSpace Independent
  Publishing Platform, North Charleston (South Carolina), US, 2012).

\bibitem{Oli02}
A.~G. Oliveira and R.~de~Abreu, {\em arXiv.org} astro-ph/0208365 (2002).

\bibitem{Oli21}
A.~G. Oliveira, Physics for expanding space --- i. the phenomenon  (2021),
  manuscript (priv. comm.).

\bibitem{Hey14}
Y.~Heymann, {\em Progress in Physics} {\bf 10(3)} (2014) 4.

\bibitem{And88}
T.~B. Andrews, {\em Physica B+C} {\bf 151}  (1988) 351.

\bibitem{And98}
T.~B. Andrews, Derivation of newton’s law of gravitation and discovery of the
  unique normal modes of the universe, in {\em Causality and Locality in Modern
  Physics\/},  eds. G.~Hunter, S.~Jeffers and J.-P. Vigier (Springer,
  Dordrecht, 1998) p. 135.

\bibitem{Haw60}
G.~S. Hawkins, {\em Astron. J.} {\bf 65}  (1960)  ~52.

\bibitem{Haw62a}
G.~S. Hawkins, {\em Nuovo Cimento} {\bf 23}  (1962) 1021.

\bibitem{Haw62b}
G.~S. Hawkins, {\em Nature} {\bf 194}  (1962) 563.

\bibitem{Haw93}
G.~S. Hawkins, Eternal universe, in {\em Encyclopedia of Cosmology\/},  ed.
  N.~S. Hetherington (Garland Publishing Inc, New York, 1993) p. 196.

\bibitem{Bar81}
Y.~V. Baryshev, {\em Astrofizicheskie Issledovaniia Izvestiya Spetsial'noj
  Astrofizicheskoj Observatorii} {\bf 14}  (1981)  ~24.

\bibitem{Gho97}
A.~Ghosh, {\em J. Astrophys. Astron.} {\bf 18}  (1997) 449.

\bibitem{Bro01}
H.~Broberg, The geometry of acceleration in space-time: Application to the
  gravitational field and particles, in {\em Gravitation, Electromagnetism and
  Cosmology: Toward a New Synthesis\/},  ed. K.~Rudnicki (Apeiron, Montreal,
  2001) p.~9.

\bibitem{Cra06}
D.~F. Crawford, {\em Curvature Cosmology} (BrownWalker Press, Boca Raton
  (Florida), US, 2006).

\bibitem{Iva06}
M.~A. Ivanov, Low-energy quantum gravity leads to another picture of the
  universe, in {\em 1st Crisis in Cosmology Conference (AIP Conf. Ser.
  822(1))\/},  eds. E.~J. Lerner and J.~B. Almeida (AIP, Melville, 2006) p.
  187.

\bibitem{Fis10}
E.~Fischer, {\em Astrophys. Space Sci.} {\bf 325}  (2010) 69.

\bibitem{Mic16}
M.~Michelini, {\em Applied Physics Research} {\bf 8}  (2016) 19.

\bibitem{Ros06}
D.~Roscoe, {\em Apeiron} {\bf 13}  (2006) 206.

\bibitem{Spa21}
A.~D. A.~M. Spallicci, J.~A. Helay\"el-Neto, M.~L\'opez-Corredoira and
  S.~Capozziello, {\em Eur. Phys. C} {\bf 81}  (2021) id. 4.

\bibitem{Alf06}
A.~Alfonso-Faus, Mass-boom versus big-bang: An alternative model, in {\em 1st
  Crisis in Cosmology Conference (AIP Conf. Ser. 822(1))\/},  eds. E.~J. Lerner
  and J.~B. Almeida (AIP, Melville, 2006) p. 107.

\bibitem{Sht93}
E.~I. Shtyrkov, A new interpretation of cosmological redshifts: Variable light
  velocity, in {\em Progress in New Cosmologies\/},  eds. A.~H. C., K.~C. R.
  and R.~K. (Springer, Boston, MA, 1993) p. 327.

\bibitem{Hun10}
J.~Hunter, {\em J. Cosmol.} {\bf 6}  (2010) 1485.

\bibitem{Gar01}
V.~I. Garaimov, Time and entropy, in {\em The emergence of cosmic structure
  (AIP Conf. Proc. 666)\/},  eds. S.~S. Holt and C.~S. Reynolds (AIP, Melville,
  2003) p. 361.

\bibitem{Tag08}
I.~N. Taganov, {\em Quantum cosmology: deceleration of time} (TIN,
  St.-Petersburg, 2008).

\bibitem{Urb08}
K.~Urbanowski, On a possible quantum contribution to the red shift, in {\em
  Practical Cosmology, vol. 1\/},  eds. Y.~Baryshev, I.~N. Taganov and
  P.~Teerikorpi (TIN, St.-Petersburg, 2008) p. 117.

\bibitem{Hoy64}
F.~Hoyle and J.~V. Narlikar, {\em Proc. R. Soc. London} {\bf A282}  (1964) 191.

\bibitem{Nar77}
J.~V. Narlikar, {\em Ann. Phys.} {\bf 107}  (1977) 325.

\bibitem{Nar93}
J.~V. Narlikar and H.~C. Arp, {\em Astrophys. J.} {\bf 405}  (1993) 51.

\bibitem{Bar02b}
G.~Barber, {\em arXiv.org} gr-qc/0212111 (2002).

\bibitem{Bar02a}
G.~Barber, {\em Astrophys. Space Sci.} {\bf 282}  (2002) 683.

\bibitem{Bar03b}
G.~Barber, {\em arXiv.org} gr-qc/0302088 (2003).

\bibitem{Bar03a}
G.~Barber, {\em Astrophys. Space Sci.} {\bf 305}  (2006) 169.

\bibitem{Ran14}
C.~Ranzan, {\em Am. J. Astron. Astrophys.} {\bf 2(5)} (2014) 47.

\bibitem{Ran15}
C.~Ranzan, {\em Physics Essays} {\bf 28(4)} (2015) 455.

\bibitem{Ein17}
A.~Einstein, Kosmologische Betrachtungen zur allgemeinen Relativit\"atstheorie,
  in {\em Sitzungsberichte der K\"oniglich Preussischen Akademie der Wissenschaften\/}
  (Akademie der Wissenschaften, Berlin, 1917) p. 142.

\bibitem{Boe07}
C.~G. Boehmer, L.~Hollenstein and F.~S.~N. Lobo, {\em Phys. Rev. D} {\bf 76}
  (2007) id. 084005.

\bibitem{Van84}
T.~Van~Flandern, Is the gravitational constant changing?, in {\em Precision
  Measurements and Fundamental Constants II},  eds. B.~N. Taylor and W.~D.
  Phillips (National Bureau of Standards Special Publication 617, 1984, Gaithersburg) p.
  625.

\bibitem{Tro87}
V.~S. Troitskii, {\em Astrophys. Space Sci.} {\bf 139}  (1987) 389.

\bibitem{Van93}
T.~Van~Flandern, {\em Dark Matter, Missing Planets and New Comets} (North
  Atlantic Books, Berkeley, 1993).

\bibitem{Fey95}
R.~P. Feynman, F.~B. Morinigo and W.~G. Wagner, {\em Feynman lectures on
  gravitation} (Addison-Wesley, Reading, MA, 1995).

\bibitem{Bar08b}
Y.~V. Baryshev, Field fractal cosmological model as an example of practical
  cosmology approach, in {\em Practical Cosmology 1\/},  eds. Y.~V. Baryshev,
  I.~N. Taganov and P.~Teerikorpi (TIN, St.-Petersburg, 2008) p. 60.

\bibitem{Bon61}
H.~Bondi, {\em Cosmology}, 2nd edn. (Cambridge Univ. Press, Cambridge, 1961).

\bibitem{Ner21}
W.~von Nernst, {\em The Structure of the Universe in Light of our Research}
  (Jules Springer, Berlin, 1921).

\bibitem{Lop13b} M. L\'opez-Corredoira, 
{\em Int. J. Mod. Phys. D} {\bf 22} (2013) 1350032.

\bibitem{Lop17} M. L\'opez-Corredoira, C. M. Guti\'errez and R. T. G\'enova-Santos,
{\em Astrophys. J.} {\bf 840} (2017) 62.

\bibitem{Yan16}
J.~L. Yang, {\em Int. J. Advanced Res. Phys. Sci.} {\bf 3}  (2016) 5.


\end{thebibliography}



\end{document}